\documentclass[preprint]{elsarticle}

\usepackage[utf8]{inputenc}
\usepackage{todonotes}
\usepackage{url}
\usepackage{amsmath}
\usepackage{amssymb}
\usepackage{xspace}
\usepackage[capitalize]{cleveref}
\usepackage{siunitx}
\usepackage{pdflscape}
\usepackage{chemformula}
\usepackage{multirow}
\usepackage{subcaption}
\usepackage[section]{placeins}
\usepackage{lineno}

\numberwithin{equation}{section}

\graphicspath{{./img/}}

\renewcommand{\div}{\nabla\cdot\!}
\newcommand{\grad}{\nabla\!}
\newcommand{\laplace}{\nabla^2}
\newcommand\numberthis{\addtocounter{equation}{1}\tag{\theequation}}

\begin{document}

\title{Modeling tissue perfusion in terms of 1d-3d embedded mixed-dimension coupled problems with distributed sources}
\author[1]{Timo Koch\corref{cor1}}
\ead{timo.koch@iws.uni-stuttgart.de}
\author[1]{Martin Schneider}
\author[1]{Rainer Helmig}
\author[3]{Patrick Jenny}

\cortext[cor1]{Corresponding author}
\address[1]{Department of Hydromechanics and Modelling of Hydrosystems, University of Stuttgart, Pfaffenwaldring 61, 70569 Stuttgart, Germany}
\address[3]{Institute for Fluid Dynamics, ETH Zürich, Sonneggstrasse 3, 8092 Zürich, Switzerland}

\begin{abstract}
We present a new method for modeling tissue perfusion on the capillary scale. The microvasculature is represented by a network of one-dimensional vessel segments embedded in the extra-vascular space. Vascular and extra-vascular space exchange fluid over the vessel walls. This exchange is modeled by distributed sources using smooth kernel functions for the extra-vascular domain. It is shown that the proposed method may significantly improve the approximation of the exchange flux, in comparison with existing methods for mixed-dimension embedded problems.
Furthermore, the method exhibits better convergence rates of the relevant quantities due to the increased regularity of the extra-vascular pressure solution.
Numerical experiments with a vascular network from the rat cortex show that the error in the approximation of the exchange flux for coarse grid resolution may
be decreased by a factor of $3$. This may open the way for computing on larger network domains,
where a fine grid resolution cannot be achieved in practical simulations due to constraints in computational resources,
for example in the context of uncertainty quantification.
\end{abstract}


\maketitle

\section{Introduction}

Most biological tissues contain embedded blood vessels that supply the cells in the extra-vascular space with oxygen and nutrients.
Mass transfer to the extra-vascular space is essential for, e.g., the function of the liver, the clearance of substances in the blood by the kidneys, or the proliferation of therapeutic agents. Moreover, fluid exchange as well as transport of molecules over the capillary walls are of particular importance, if the blood-brain-barrier (\textit{BBB}) is impaired, e.g. due to brain diseases like multiple sclerosis~\citep{minagar2003blood,kermode1990breakdown,Verma2017,Koch2018b} or glioblastoma~\citep{Verma2017,Watkins2014bbb}. The accurate description of the fluid exchange is essential to analyze, e.g., advanced MR imaging techniques for the brain~\citep{tofts1991measurement,boxerman2006relative}. Theoretical models can help to understand and quantify the relevant flow and transport processes in biological tissue~\citep{heye2016tracer,Koch2018b,Holter2018}.

Due to the large number of vessels in the microvasculature, fully three-dimensionally resolved models are only feasible for small tissue samples.
Herein, we consider a class of models, for which the microvasculature is represented by a network of vessel segments embedded in the extra-vascular space, and the flow in the vessels is modeled by one-dimensional partial differential equations (\textit{PDE}s), whereas the extra-vascular space is modeled as a homogenized porous medium by a three-dimensional PDE. The PDEs are coupled via source terms. Such models have been used to study the proliferation of cancer drugs~\cite{cattaneo2014computational,d2007multiscale,shipley2010multiscale}, nano-particle transport in hypothermia therapy~\cite{Nabil2015}, oxygen proliferation~\cite{Linninger2013,Fang2008,chapman2008blood,secomb2004green,secomb2000theoretical}, and contrast agent perfusion~\cite{Holter2018,Koch2018b}.

This work focuses on the application of biological tissue perfusion. However, the presented method can also be applied to model
other network-like structured embedded in porous media, such as plant root systems~\cite{Doussan1998,Javaux2008,Koch2018a} and
wells in petroleum engineering or geothermal applications~\cite{Cerroni2019,Wolfsteiner2003,Aavatsmark2003,alkhoury2005}.

\subsection{Mixed-dimension embedded model with line sources}
\label{sec:mixeddim_intro}

\begin{figure}
 \centering
 \includegraphics[width=0.4\textwidth]{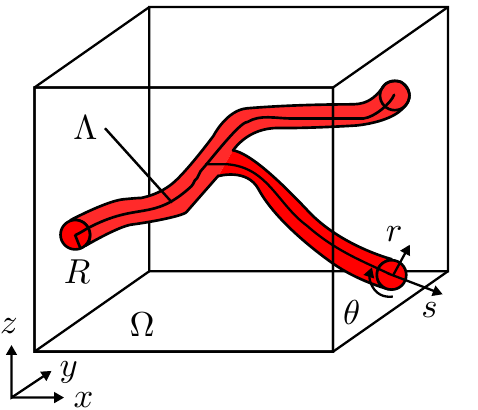}
 \caption{A capillary bifurcation with center-line $\Lambda$ embedded in the extra-vascular space $\Omega$.
 A segment-local cylindrical coordinate system is shown with the coordinates $r$, $\theta$, $s$, as well as a Cartesian coordinate system for $\mathbb{R}^3$ with the coordinates $x$, $y$, $z$. The vessel radius at any segment cross-section is denoted by $R$.}
 \label{fig:domains}
\end{figure}
We consider a region of biological tissue, $\Omega\subset\mathbb{R}^3$, comprising cells, the extra-cellular matrix, and interstitial fluid. Embedded
in $\Omega$, a network structure of tubular vessel segments, $\Lambda\subset\mathbb{R}^3$, models the microvasculature. The domains are conceptually shown in \cref{fig:domains}.
Blood is assumed to behave like a Newtonian fluid with constant density, $\rho_B$, and dynamic viscosity, $\mu_B$.
Blood flow through the vessel lumen is governed by the Hagen-Poiseuille law. These strong assumptions
merely simplify the following analysis, but are not required for the derivation of the presented method. The extra-vascular space is
modeled as a porous medium with porosity $\phi$ and isotropic intrinsic permeability $\kappa$. The interstitial fluid flow is
described by Darcy's law. The two compartments exchange mass over the vessel walls, $\mathbb{W}$. This transmural mass exchange is commonly
modeled by Starling's law~\citep{Starling1896} as
\begin{equation}
\hat{q}_m = \rho_I L_p S \left[ (\tilde{p}_v - \tilde{p}_{t,\mathbb{W}} - \sigma\Delta\Pi \right],
\end{equation}
where transmural mass flux, $\hat{q}_m$ (\si{\kg\per\s\m\squared}), is formulated in terms of the filtration coefficient, $L_p$ (\si{\m\per\pascal\s}), the cross-sectional vessel wall surface $S = 2\pi R$, the mean hydraulic pressures in the lumen, $\tilde{p}_v$, and on the outer vessel wall surface, $\tilde{p}_{t,\mathbb{W}}$, the oncotic pressure difference, $\Delta\Pi = \Pi_v - \Pi_t$, and the osmotic reflection coefficient $\sigma \in \left[ 0,1 \right]$.
The oncotic pressure is exerted by large proteins in the blood plasma such as albumin. The difference in oncotic pressure between the vascular and the extra-vascular space effectively pulls fluid into the vessels. It can be assumed constant in the capillary bed, with $\Delta\Pi = \SI{2633}{\pascal}$~\citep{levick1991}.
Furthermore, we choose $\sigma = 1$, which corresponds to the vessel wall modeled as a perfect selectively-permeable membrane.
Thus, the problem can be reformulated in terms of effective pressures, $p_\alpha = \tilde{p}_\alpha - \Pi_\alpha, \; \alpha \in \{ v, t \}$.

Flow in the given tissue region can be described by a coupled system of partial differential equations, \citep[cf.][]{cattaneo2014computational,d2007multiscale,Koch2018b},

\begin{subequations}
\label{eq:linesources}
\begin{align}
  - \frac{\partial}{\partial s}\left( \rho_B\frac{ \pi R^4}{8\mu_B} \frac{\partial p_v}{\partial s} \right) &= -\hat{q}_m & \text{in } \Lambda, \label{eq:vessel} \\
  - \div \left( \frac{\rho_I}{\mu_I} \kappa \grad p_t \right) &= \hat{q}_m \delta_\Lambda & \text{in } \Omega,\label{eq:tissue} \\
  \hat{q}_m &= \rho_I L_p S (p_v - p_{t,\mathbb{W}}), \label{eq:exchange}
\end{align}
\end{subequations}
where $s$ is the local coordinate in axial direction of the vessel segments, $R$ denotes the equivalent vessel radius corresponding to the cross-sectional area of the vessel, $A_v = \pi R^2$, and $\rho_I$ and $\mu_I$ are density and dynamic viscosity of the interstitial fluid, respectively.
The Dirac delta function $\delta_\Lambda$ with the property
\begin{equation}
\int_\Omega f\delta_\Lambda \text{d}x = \int_\Lambda f \text{d}s \quad \forall f
\end{equation}
restricts $\hat{q}_m$ in \cref{eq:tissue} on the vessel center-line. The derivation of the formulation with line sources, \cref{eq:linesources}, is discussed in~\citep{d2007multiscale} and based on scaling the vessel radius to the zero limit under the constraint of flux equality.
However, solutions to \cref{eq:linesources} exhibit a singularity at $r = 0$, which is both nonphysical and challenging for numerical methods. The author of \citep{d2007multiscale,DAngelo2012} proves convergence of standard finite element methods in weighted norms. The authors of \citep{koeppl2016} can show optimal convergence up to a log-factor in classical but local norms excluding a small neighborhood around the singularity. It is argued that the solution is only physically meaningful for $r>R$. However, numerical experiments with standard finite element methods show that optimal convergence rates can only be achieved, if the cell size $h_\Omega$ of the grid discretizing $\Omega$, is in the order of magnitude of the vessel radius $R$ and smaller.
In \citep{secomb2004green}, Green's functions are used to solve a system of equations similar to  \cref{eq:linesources}. The authors exploit the superposition property of the Laplace operator. The vessel network is represented by a collection of point sources along the vessel center-lines. The solution in $\Omega$ is constructed by adding up the contributions of all point sources and a smooth correction function to satisfy the boundary conditions. However, due to global interaction of those contributions, the numerical solution of the point source strengths results in dense system matrices. In \citep{Gjerde2018}, the support of the line source contributions is narrowed by a smooth cut-off function and the correction function is numerically approximated, yielding sparse system matrices. This method is also known as subtraction method from the field of electroencephalography (EEG) source reconstruction~\citep{Engwer2017}. A local version of such methods has been well-studied in petroleum engineering and is known as the Peaceman well model~\citep{Peaceman1983,chen2009well} in its simplest form. Local approximations are difficult to construct for arbitrarily oriented wells~\citep{Wolfsteiner2003,Aavatsmark2003}. Moreover, these well models are derived in a discrete setting.

\subsection{Other mixed-dimension embedded models}

A different approach is taken in \citep{koeppl2018}. Instead of using line sources, the exchange term between vessels and extra-vascular space is considered on the actual surface of the cylindrical vessel tubes, thus increasing the dimension of the source term by one. As a consequence, the singularity is replaced by a smoother continuation of the pressure function for $r < R$, leading to better numerical properties. However, good approximations of the exchange flux $\hat{q}_m$ still requires very fine grid resolutions.
There is the necessity for a method with a good approximation of the exchange term, even if a very fine grid resolution is not computationally feasible.
Moreover, we are seeking a method that is robust and efficient regarding large local variations of vessel radii, which readily occur in complex vessel networks.

\subsection{Scope of this work}

In the following, we discuss a different formulation of \cref{eq:linesources}, where the source term in the extra-vascular space is approximated as a distribution over a three-dimensional neighborhood of the vessel using a smoothing kernel. This novel approach for modeling tissue perfusion, is introduced in a continuous setting in~\cref{sec:method}. The model is discretized using a cell-centered finite volume method in \cref{sec:disc}. The robustness of the method and a comparison with existing methods is demonstrated and discussed in a series of numerical experiments in~\cref{sec:numerical}.

\section{Mixed-dimension embedded model with distributed sources}
\label{sec:method}

Here, we consider the same set of equations as in~\cref{sec:mixeddim_intro} with a modified source term,
\begin{subequations}
\label{eq:kernel}
\begin{align}
  - \frac{\partial}{\partial s}\left( \rho_B\frac{ \pi R^4}{8\mu_B} \frac{\partial p_v}{\partial s} \right) &= -\hat{q}_m & \text{in } \Lambda, \label{eq:vessel2} \\
  - \div \left( \frac{\rho_I}{\mu_I} \kappa \grad p_t \right) &= \hat{q}_m \Phi_\Lambda & \text{in } \Omega,\label{eq:tissue2} \\
  \hat{q}_m &= \rho_I L_p S (p_v - p_{t,0})\Xi, \label{eq:exchange2}
\end{align}
\end{subequations}
where $\Phi_\Lambda$ is a set of kernel functions $\Phi_{\Lambda_i}$ that distribute $\hat{q}_m$ over a small tubular support region, $\mathcal{S}(\Phi_{\Lambda_i})$, with radius $\varrho(s)$, around a vessel segment $i$, such that $\Phi_{\Lambda_i} = 0$ outside the support region. The flux $\hat{q}_m$ in \cref{eq:exchange2} is now formulated in terms of the extra-vascular pressure $p_{t,0}$, evaluated at the vessel center-line. The function $\Xi = \Xi(\varrho(s), \Phi(s), \cdots)$ is a flux scaling factor depending on the support radius, and the chosen kernel function. We choose kernel functions $\Phi_{\Lambda_i}(s)$ with the property
\begin{equation}
\int\displaylimits_0^{l_i}\!\int\displaylimits_0^{2\pi}\!\int\displaylimits_0^{\varrho(s)} \Phi_{\Lambda_i} r\;\text{d}r\text{d}\theta\text{d}s = l_i,
\end{equation}
where $r$, $\theta$, $s$ are the radial, angular, and axial coordinate in a segment-local cylinder coordinate system, and $l_i$ is the length of segment $i$.
On close observation, one can see that if the scaling factor is chosen as
\begin{equation}
\Xi = \frac{(p_v - p_{t,\mathbb{W}})}{(p_v - p_{t,0})},
\label{eq:xi}
\end{equation}
then in the limit of $\varrho \rightarrow 0$, \cref{eq:kernel} is equal to \cref{eq:linesources}. Hence, the mixed-dimension embedded method with line sources is a special case of the presented method, where $\Phi_\Lambda := \delta_\Lambda$.
The formulation of \citep{koeppl2018} is obtained by choosing kernel functions that are $S^{-1} = (2\pi R)^{-1}$ on the vessel surface, and zero elsewhere.
A similar formulation to \cref{eq:kernel} is introduced in \citep{Karvounis2016} for an embedded discrete fracture model (2d-3d), but not further analyzed or numerically exploited.

\subsection{Flux scaling factor $\Xi$ (single straight vessel)}
\label{sec:xi}

\begin{figure}
\centering
\includegraphics[width=0.9\textwidth]{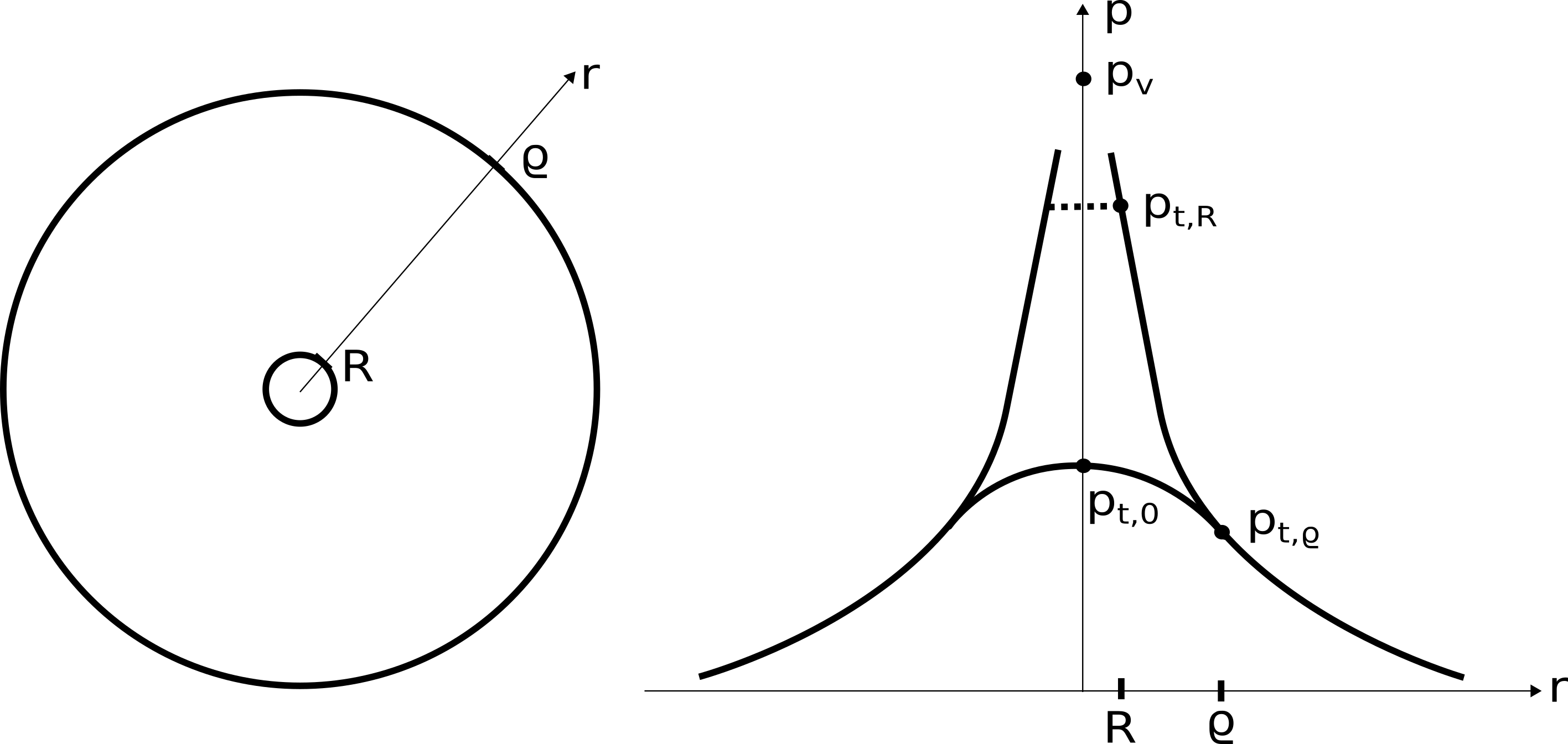}
\caption{Location of different pressures in radial distance to the vessel center-line, as well as in relation to each other, for the case of extravasion. The singular solution approaches $\infty$ for $r\rightarrow 0$, while the regularized solution has the finite value $p_{t,0}$ at $r=0$.}
\label{fig:solution}
\end{figure}
For non-zero support, $\varrho > 0$, smooth kernel functions regularize the pressure solution in the kernel support region. Hence, the solutions to \cref{eq:kernel} have no singularities. The scaling factor $\Xi$ can be analytically derived in terms of the kernel support radius $\varrho$ and the constraint~\cref{eq:xi} under some assumptions. First, let us consider a single straight and long vessel. Looking at a cross-sectional plane cutting through this vessel (see \cref{fig:solution}), we assume that the flow field in a small neighborhood around the vessel is strictly radial, and can thus be considered two-dimensional.
Now, let us exemplarily consider the following radial kernel function
\begin{equation}
  \Phi^\text{const}(r) = \begin{cases}
    \frac{1}{\pi\varrho^2} & r \leq \varrho,\\
    0 & r > \varrho. \end{cases}
\end{equation}
The flux outside the kernel support region, $r>\varrho$, is identical to the flux of the formulation with line sources,
as the same total mass is injected in both formulations if the flux scaling factor $\Xi$ is chosen as in \cref{eq:xi}.
The pressure solution has to satisfy
\begin{align}
\label{eq:laplace}
- \frac{1}{r}\frac{\partial}{\partial r}\left( r\frac{\partial p_t}{\partial r} \right) &= \frac{\mu_I}{\rho_I \kappa}\hat{q}_m\Phi^\text{const}(r).
\end{align}
The pressure outside the kernel support region admits the following solution
\begin{equation}
  p_t(r) = p_{t,\mathbb{W}} - \frac{\hat{q}_m \mu_I}{2\pi\rho_I \kappa}\ln\left(\frac{r}{R}\right), \quad r > \varrho, \label{eq:singularsol}
\end{equation}
derived from the fundamental solution of the Laplace equation. Note that the mean pressure on the vessel surface, $p_{t,\mathbb{W}}$, is equal to the pressure
evaluated at $r=R$, if $\varrho \leq R$. To derive the pressure solution inside the kernel support, we apply Leibniz's rule to~\cref{eq:laplace}
\begin{align}
\frac{\partial p_t}{\partial r} &= -\frac{1}{r} \int_0^r \frac{\mu_I}{\rho_I \kappa}\hat{q}_m\Phi^\text{const}(r') r' \text{d}r',
\end{align}
and integrating once more yields
\begin{equation}
p_t(r) = -\frac{\hat{q}_m \mu_I}{2\pi\rho_I \kappa}\frac{r^2}{2\varrho^2} + C_0, \quad r \leq \varrho.
\label{eq:kernelc0}
\end{equation}
It is easy to verify that for $r=\varrho$, \cref{eq:singularsol,eq:kernelc0} have equal derivatives which ensures flux continuity.
The integration constant $C_0$ is determined such that the pressure is continuous at $r=\varrho$, yielding
\begin{equation}
p_t(r) = \begin{cases}
p_{t,\mathbb{W}} -\frac{\hat{q}_m \mu_I}{2\pi\rho_I \kappa}\left[ \frac{r^2}{2\varrho^2} + \ln\left(\frac{\varrho}{R}\right) - \frac{1}{2} \right] & r \leq \varrho,\\
  p_{t,\mathbb{W}} - \frac{\hat{q}_m \mu_I}{2\pi\rho_I \kappa}\ln\left(\frac{r}{R}\right) & r > \varrho.
   \end{cases}
\label{eq:ana_constkernel}
\end{equation}
This function is shown qualitatively in \cref{fig:solution}.
To derive a suitable $\Xi$, let us first evaluate \cref{eq:ana_constkernel} at $r=0$, so that $p_{t,\mathbb{W}}$ is expressed in terms of $p_{t,0}$,
\begin{equation}
p_{t,0} = p_{t,\mathbb{W}} -\frac{L_p R \mu_I}{\kappa}(p_v - p_{t,0})\Xi\left[ \ln\left(\frac{\varrho}{R}\right) - \frac{1}{2} \right],
\label{eq:pt0andptw}
\end{equation}
where $\hat{q}_m$ was replaced by inserting \cref{eq:exchange2}.
It directly follows from \cref{eq:pt0andptw,eq:xi} that the flux scaling factor can be expressed independently of the pressure,
\begin{align}
\Xi &= \frac{(p_v - p_{t,\mathbb{W}})}{(p_v - p_{t,0})} = \frac{(p_v - p_{t,0}) - \frac{L_p R \mu_I}{\kappa}(p_v - p_{t,0})\Xi\left[ \ln\left(\frac{\varrho}{R}\right) - \frac{1}{2} \right]}{(p_v - p_{t,0})} \\
 &= \left\{ 1 + \frac{R L_p\mu_I}{\kappa}\left[\ln\left(\frac{\varrho}{R}\right)-\frac{1}{2} \right]\right\}^{-1} \quad (\text{for} \quad \Phi^\text{const}).
\label{eq:xi_const}
\end{align}
The flux factor has a very similar structure, when derived for other kernel functions. For instance, for the cubic kernel function
\begin{equation}
  \Phi^\text{cubic}(r) = \begin{cases}
    \frac{10}{3\pi\varrho^2}\left[\frac{2r^3}{\varrho^3} - \frac{3 r^2}{\varrho^2} + 1\right] & r \leq \varrho,\\ 0 & r > \varrho, \end{cases}
\end{equation}
we obtain
\begin{equation}
p_t(r) = \begin{cases}
p_{t,\mathbb{W}} -\frac{\hat{q}_m \mu_I}{2\pi\rho_I \kappa}\left[ \frac{r^2}{\varrho^2}\left( \frac{8r^3}{15\varrho^3} - \frac{5r^2}{4\varrho^2} + \frac{5}{3}  \right) + \ln\left(\frac{\varrho}{R}\right) - \frac{19}{20} \right] & r \leq \varrho,\\
  p_{t,\mathbb{W}} - \frac{\hat{q}_m \mu_I}{2\pi\rho_I \kappa}\ln\left(\frac{r}{R}\right) & r > \varrho,
   \end{cases}
\end{equation}
and
\begin{equation}
\Xi(\Phi^\text{cubic}) = \left\{ 1 + \frac{R L_p\mu_I}{\kappa}\left[ \ln\left(\frac{\varrho}{R}\right)-\frac{19}{20} \right]\right\}^{-1}.
\label{eq:xi_cubic}
\end{equation}
The dimensionless group $\Theta = R L_p\mu_I \kappa^{-1}$ is the ratio of the hydraulic conductivity of the vessel wall to the hydraulic conductivity of the extra-vascular space. If the filtration coefficient $L_p$ is low relative to $\kappa$, the regularized pressure profile is rather flat, so that the difference between $p_{t,R}$ and $p_{t,0}$ is low. This is reflected in a flux scaling factor close to $1$. If the filtration coefficient is elevated, e.g. when the BBB is impaired in a tumor, $\Theta$ can be larger than $1$, hence $\Xi$ may significantly defer from $1$. Comparing \cref{eq:xi_cubic,eq:xi_const}, we observe that a kernel function with higher weights towards the vessel center-line tends to result in a $\Xi$ further away from $1$.

\subsection{Kernel support radius $\varrho$}

Considering that the physically meaningful part of the pressure solution is actually given for $r \geq R$, if $\varrho < R$, our formulation gives identical solutions (for $r \geq R$) to the methods discussed in \citep{d2007multiscale,koeppl2018}. However, the smoothness of the solution for $r < R$ can be controlled by the choice of kernel functions. For $\varrho > R$, the physical pressure solution $p_t$ is altered. However, the source term as well as $p_v$, remain the same for the derived $\Xi$.
For $\varrho > R$, the action of the coupling term on $p_t$ is distributed over an artificially enlarged vessel volume.

\subsection{Multiple vessels}
\label{sec:theory_multiple}

In the capillary bed, vessels form a dense network. The vessel diameters are small, such that average distances between vessels are about one order of magnitude larger than the radii, and the vessel volume fraction is in the range of \SIrange{2}{5}{\percent}. In the following, we present our considerations for choosing $\Xi$ in such a network.

To this end, let us consider a setup of $N$ parallel long vessels with given boundary conditions, such that the resulting pressure solution in each plane, $\mathbb{P} \in \mathbb{R}^3$, perpendicular to the vessel is equal to the two-dimensional problem on that plane (the in-plane solutions are independent of the solution in any parallel plane).
Due to the linearity of the Laplace operator the solution can be split into contributions by the individual vessels (superposition principle),
\begin{equation}
p_t(\boldsymbol{x}) = \sum\limits_{i=1}^N p_{t,i}(\boldsymbol{x}),
\label{eq:superpos}
\end{equation}
where $r_i = \vert\vert \boldsymbol{x}_i - \boldsymbol{x} \vert\vert_2$ is the distance between a point $\boldsymbol{x} \in \mathbb{P}$ and the center $\boldsymbol{x}_i$ of the vessel $i$.
The source term contributions of the individual vessels are denoted by $\hat{q}_{m,i}$. Note that it depends on $p_t$ rather than $p_{t,i}$,
\begin{equation}
\hat{q}_{m,i} = 2\pi R_i \rho_I L_p (p_v - p_t(\boldsymbol{x}_i))\Xi_i,
\end{equation}
so that the partial solutions $p_{t,i}$ are not independent of each other. Each partial solution, $p_{t,i}$, assumes the same form as \cref{eq:ana_constkernel}.
Furthermore, $p_{t,i}$ is a harmonic function, satisfying the Laplace equation $\div\grad p_{t,i} = 0$, for $r_i > \varrho_i$.
Assuming, that the kernel support regions of two neighboring vessels do not overlap, we observe that
\begin{equation}
p_{t,j}(\boldsymbol{x}_i) = \frac{1}{2\pi}\int_0^{2\pi} \left. p_{t,j} \right\rvert_{R_i} \text{d}\theta =: p_{t,j,\mathbb{W}_i} =: C_{j,i}
\label{eq:harmonicassumption}
\end{equation}
using the mean value property of harmonic functions~\citep[][p.4f]{axler1992harmonicfunctions},
where $p_{t,j,\mathbb{W}_i}$ is defined as the average of $p_{t,j}$ over the cross-sectional vessel surface of vessel $i$.
From \cref{eq:superpos,eq:harmonicassumption} follows that
\begin{equation}
p_{t,i}(\boldsymbol{x}_i) - p_{t,i,\mathbb{W}_i} =
p_{t,i}(\boldsymbol{x}_i) + \sum\limits_{\substack{j=1 \\ j \neq i}}^N C_{j,i} - (p_{t,i,\mathbb{W}_i}  + \sum\limits_{\substack{j=1 \\ j \neq i}}^N C_{j,i})
= p_{t}(\boldsymbol{x}_i) - p_{t,\mathbb{W}_i},
\end{equation}
where
\begin{equation}
\quad p_{t,\mathbb{W}_i} := \frac{1}{2\pi}\int_0^{2\pi} \left. p_{t} \right\rvert_{R_i} \text{d}\theta.
\end{equation}
This allows us to derive the flux scaling factors $\Xi_i$ analogously to the single vessel case (cf. \cref{eq:pt0andptw,eq:xi_const}),
\begin{equation}
\Xi_i(\Phi^\text{const}) = \frac{p_{v,i} - p_{t,\mathbb{W}_i}}{p_{v,i} - p_{t}(\boldsymbol{x}_i)} = \left\{ 1 + \frac{R_i L_p\mu_I}{\kappa}\left[\ln\left(\frac{\varrho_i}{R_i}\right)-\frac{1}{2} \right]\right\}^{-1}.
\end{equation}
In \cref{sec:numerical_multiple}, we construct an analytical solution with multiple parallel vessels using these observations.

Unfortunately, the derivation using the mean value property of harmonic functions is no longer valid for arbitrarily-oriented vessels.
In a fully three-dimensional setup, the mean value property concerns the integral over a sphere, while in our model we are integrating over the boundary of a circular cross-section.
However, numerical experiments have shown that using an unmodified flux scaling factor does not introduce a significant model error.
A possible explanation is that at a large enough distance to the neighboring vessels
the integral over the circular cross-section is expected to be very close to the integral over a sphere with the same radius and center point,
given that the pressure gradient decays with $r_i^{-1}$.
Moreover, realistic segmented vessel networks contain vessel bifurcations and may contain bends with sharp angles.
In the vicinity of such features, the kernel support regions
of two connected vessels may overlap. In such a case the correct flux scaling factor is not easily determined.
However, we herein still employ the flux scaling factor shown in~\cref{eq:xi_const}, assuming
that possible errors are small, only occur in the close vicinity of such features and that the influence on the global solution is negligible in realistic applications.
These assumptions and the corresponding modeling errors are investigated in a numerical experiment in \cref{sec:numerical_network}.

\section{Discretization}
\label{sec:disc}

\begin{figure}
 \centering
 \includegraphics[width=\textwidth]{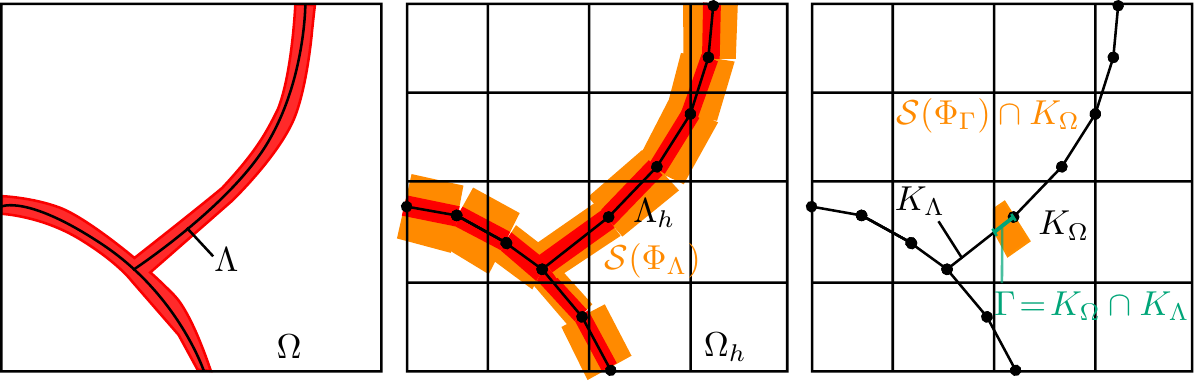}
 \caption{Visualization of the discretization. The domains $\Omega$ and $\Lambda$ (left) are represented by a set of control volumes $K_\Omega \in \Omega_h$ and $K_\Lambda \in \Lambda_h$ (middle). The red area shows a tube with a radius $R$ around the vessel center-line. The orange area visualizes the kernel support $\mathcal{S}(\Phi_{\Lambda_h})$ of the kernel functions $\Phi_{\Lambda_h}$ in the middle, and on the right, the kernel support associated with the intersection $\Gamma$ intersected with the control volume $K_\Omega$. In this example, the kernel functions with cylindrical support, are chosen to have radii $\varrho > R$.}
 \label{fig:disc}
\end{figure}
We discretize \cref{eq:kernel} using a cell-centered finite volume method with two-point flux approximation. To this end,
$\Omega$ and $\Lambda$ are decomposed into two independent meshes $\Omega_h$ and $\Lambda_h$ consisting of control volumes $K_\Omega\in\Omega_h$ and $K_\Lambda\in\Lambda_h$. Herein, $K_\Omega$ are chosen to be hexahedra and
$K_\Lambda$ are line segments. Furthermore, the control volume boundary, $\partial K$, can be split into a finite number of faces $\sigma\subset\partial K$, such that $\sigma = K\cap L$, given a neighboring control volume $L$. Integrating \cref{eq:tissue2} over a control volume $K_\Omega$ and applying the Gauss divergence theorem to the left hand side yields
\begin{equation}
\label{eq:disc_integral}
-\int_{\partial K_\Omega}\! \left[ \frac{\rho_I}{\mu_I} \kappa \grad p_t \right]\cdot\boldsymbol{n}_{K_\Omega,\sigma}\, \text{d}A = \int_{K_\Omega}\! \hat{q}_m \Phi_\Lambda \,\text{d}x,
\end{equation}
where $\boldsymbol{n}_{K_\Omega,\sigma}$ is the unit outward-pointing normal on face $\sigma\subset\partial K_\Omega$.
The exact fluxes are approximated by numerical fluxes
\begin{equation}
F_{K_\Omega,\sigma} \approx -\int_{\sigma}\! \left[ \frac{\rho_I}{\mu_I} \kappa \grad p_t \right]\cdot\boldsymbol{n}_{K_\Omega,\sigma}.
\end{equation}
Let $\mathcal{I}$ be the set of intersections $\Gamma = K_\Omega \cap K_\Lambda$. Furthermore, let $\Phi_\Gamma \in \Phi_\Lambda$ denote a kernel function with the support $\mathcal{S}(\Phi_\Gamma)$ associated with $\Gamma$. The discrete source term is computed as
\begin{equation}
\label{eq:qkomegakernel}
Q_{K_\Omega} = \int_{K_\Omega}\! \hat{q}_m \Phi_\Lambda \,\text{d}x = \sum_{\Phi_\Gamma \in \Phi_\Lambda} Q_\Gamma \frac{1}{\vert\Gamma\vert}\int_{K_\Omega \cap \mathcal{S}(\Phi_\Gamma)} \Phi_\Gamma \,\text{d}x,
\end{equation}
where $Q_\Gamma$ is the numerical approximation of the source term integrated over the intersection $\Gamma$,
\begin{equation}
Q_\Gamma \approx  \int_\Gamma\! \rho_I L_p S (p_v - p_{t,0})\Xi \,\text{d}s.
\end{equation}
Hence, \cref{eq:disc_integral} can be reformulated as
\begin{equation}
\sum\limits_{\sigma \subset \partial K_\Omega} F_{K_\Omega,\sigma} = Q_{K_\Omega}, \quad \forall K_\Omega \in \Omega_h.
\end{equation}
The discretization of the domains $\Omega$ and $\Lambda$ are visualized in \cref{fig:disc}, including an illustration of the kernel support region.

We mention that due to the approximation of the vessel segments as cylindrical tubes around the center-line, segments may overlap, in particular, at bifurcations and bends
with sharp angles. To the knowledge of the authors, the resulting discretization errors are neglected throughout the present literature. This is a fair assumption, given
that these overlaps are small and the corresponding discretization error is expected to be small in comparison with errors resulting from the vessel network segmentation.

\subsection{Numerical fluxes $F_{K_\Omega,\sigma}$}

We compute the numerical fluxes using a two-point flux approximation,
\begin{equation}
F_{K_\Omega,\sigma} = \frac{t_{K_\Omega,\sigma}t_{L_\Omega,\sigma}}{t_{K_\Omega,\sigma} + t_{L_\Omega,\sigma}} (p_{K_\Omega} - p_{L_\Omega})
\end{equation}
for two neighboring control volumes $K$ and $L$, with the pressure degrees of freedom $p_{K_\Omega}$ and $p_{L_\Omega}$ associated with these control volumes,
and with the transmissibilities
\begin{equation}
\label{eq:transmissib}
t_{K_\Omega,\sigma} = \vert \sigma \vert\frac{\rho_I\kappa}{\mu_I}\frac{\boldsymbol{d}_{K_\Omega,\sigma}\cdot\boldsymbol{n}_{K_\Omega,\sigma}}{\vert\vert\boldsymbol{d}_{K_\Omega,\sigma}\vert\vert^2},
\end{equation}
where $\boldsymbol{d}_{K_\Omega,\sigma} = \boldsymbol{x}_\sigma - \boldsymbol{x}_{K_\Omega}$ is a vector from the center of the control volume $K_\Omega$ to the center of the face $\sigma$,
and $\vert \sigma \vert$ denotes the area of face $\sigma$.

\subsection{Numerical fluxes $F_{K_\Lambda,\sigma}$}

Analogously to the derivation above, a discrete representation of \cref{eq:vessel2} is given by
\begin{equation}
\sum\limits_{\sigma \subset \partial K_\Lambda} F_{K_\Lambda,\sigma} = Q_{K_\Lambda}, \quad \forall K_\Lambda \in \Lambda_h,
\end{equation}
with
\begin{equation}
\label{eq:disc_f_lambda}
F_{K_\Lambda,\sigma} = t_{K_\Lambda,\sigma}(p_{K_\Lambda} - p_\sigma), \quad Q_{K_\Lambda} = \sum\limits_{\Gamma \in K_\Lambda \cap \Omega_h} Q_\Gamma,
\end{equation}
where $p_\sigma$ denotes the pressure at $x_\sigma$, and the transmissibilities are defined analogously to \cref{eq:transmissib},
\begin{equation}
t_{K_\Lambda,\sigma} = \vert \sigma \vert\frac{\pi R_K^4\rho_B}{8\mu_B}\frac{\boldsymbol{d}_{K_\Lambda,\sigma}\cdot\boldsymbol{n}_{K_\Lambda,\sigma}}{\vert\vert\boldsymbol{d}_{K_\Lambda,\sigma}\vert\vert^2},
\end{equation}
with $R_K$ denoting the vessel radius of control volume $K$.
Since $\Lambda_h$ consists of a network of segments $K_\Lambda$, it occurs that a face $\sigma$ has more than two neighboring cells, i.e. a set of neighboring cells $\mathcal{K}_\sigma\subset\Lambda_h$. At such bifurcation faces, we enforce flux conservation, just like for faces with exactly two neighbors,
\begin{equation}
\label{eq:disc_f_flux_conserved}
\sum\limits_{K_\Lambda \in \mathcal{K}_\sigma} F_{K_\Lambda,\sigma} = 0.
\end{equation}
Inserting \cref{eq:disc_f_lambda} into \cref{eq:disc_f_flux_conserved} yields
\begin{equation}
p_\sigma = \frac{\sum_{K_\Lambda \in \mathcal{K}_\sigma} t_{K_\Lambda,\sigma} p_{K_\Lambda}}{\sum_{K_\Lambda \in \mathcal{K}_\sigma} t_{K_\Lambda,\sigma}},
\end{equation}
so that the face unknown in \cref{eq:disc_f_lambda} can be replaced by an expression in terms of control volume unknowns $p_{K_\Lambda}$.

\subsection{Numerical source term $Q_\Gamma$}

In the following, we consider three different methods: the method proposed in \citep{d2007multiscale} with line sources (\textsc{ls}),
the method suggested in \citep{koeppl2018} with cylinder surface sources (\textsc{css}), and the above introduced method with distributed sources (\textsc{ds}).
The methods can be distinguished by the choice of kernel functions (see \cref{sec:method}), and the approximation of $Q_\Gamma$.
For the $\textsc{ls}$ and $\textsc{css}$ method, we are looking for the discrete approximation
\begin{equation}
Q^{\textsc{ls},\textsc{css}}_\Gamma \approx \int_\Gamma\! \rho_I L_p S \left[ p_v - \frac{1}{2\pi R}\int_{\partial\mathcal{D}_R(s)} p_t \,\text{d}\theta \right] \,\text{d}s,
\end{equation}
where $\partial\mathcal{D}_R(s)$ is the boundary of a disc at position $s$, perpendicular to the intersection segment $\Gamma$ with radius $R$. We compute
\begin{equation}
Q^{\textsc{ls},\textsc{css}}_\Gamma \approx \int_\Gamma\! \rho_I L_p S \left[ p_v - \frac{1}{2\pi R}\sum\limits_{I_K \in \mathcal{I}_{\partial\mathcal{D}_R}} \vert I_K\vert p_{t,K_\Omega}  \right] \,\text{d}s,
\end{equation}
where $\mathcal{I}_{\partial\mathcal{D}_R}$ is the set of intersections $I_K = \partial\mathcal{D}_R(s)\cap K_\Omega$, $K_\Omega \in \Omega_h$,
and the integral is approximated by a Gaussian quadrature rule.
For the method $\textsc{ds}$ we seek
\begin{equation}
\label{eq:qds}
Q^{\textsc{ds}}_\Gamma \approx \int_\Gamma\! \rho_I L_p S \left[ p_v - p_{t,0}(s) \right]\Xi \,\text{d}s.
\end{equation}
The center-line pressure $p_{t,0}(s)$ could be approximated by the pressure $p_{K_\Omega}$ in the control volume $K_\Omega$ containing $\Gamma$. However, this approximation is poor, if $K_\Omega$ is not significantly smaller than the vessel radius $R$. Assuming radial flux in a small neighborhood of $\Gamma$, we can reformulate \cref{eq:qds} as
\begin{equation}
\label{eq:reform_qds}
Q^{\textsc{ds}}_\Gamma \approx \int_\Gamma\! \rho_I L_p S \left[ p_v - p_{t,0}(s) \right]\Xi \,\text{d}s = \int_\Gamma\! \rho_I L_p S \left[ p_v - p_{t,\delta}(s) \right]\Xi_\delta \,\text{d}s,
\end{equation}
where $p_{t,\delta}$ is evaluated at some distance $\delta \leq 0$ to the vessel center-line, and $\Xi_\delta$ is a modified flux scaling factor, ensuring equality. Now, $p_{K_\Omega}$ may better approximate $p_{t,\delta}$ than $p_{t,0}$, for a certain $\delta$. As $p_{K_\Omega}$ is commonly defined as the mean pressure in the control volume $K_\Omega$, we choose $\delta$ as the mean distance of $\Gamma$ to the control volume $K_\Omega$,
\begin{equation}
\delta = \frac{1}{\vert K_\Omega \vert} \int_{K_\Omega} \min_{\boldsymbol{x'} \in \Gamma} \vert\vert \boldsymbol{x} - \boldsymbol{x'} \vert\vert_2 \,\text{d}x.
\end{equation}
The corresponding $\Xi_\delta$ is computed using the analytical derivations in \cref{sec:method}, and is dependent on the kernel functions.
Choosing $\Phi^\text{const}$, yields
\begin{equation}
\label{eq:corrected_xi}
\Xi_\delta (\Phi^\text{const}) = \left\{ 1 + \frac{R L_p\mu_I}{\kappa}\left[ \frac{\delta^2}{2 \varrho^2} + \ln\left(\frac{\varrho}{R}\right)-\frac{1}{2} \right]\right\}^{-1} \quad \text{for}\; \delta \leq \varrho,
\end{equation}
so that
\begin{equation}
Q^{\textsc{ds}}_\Gamma = \rho_I L_p S \left[ p_{K_\Lambda} - p_{K_\Omega} \right]\Xi_\delta.
\end{equation}
Denoting the maximum size of all control volumes $K_\Omega$ by $h_\Omega$, note that for $h_\Omega\rightarrow 0 \implies \delta \rightarrow 0$, and $p_{t,\delta} \rightarrow p_{t,0}$, so that this approximation is consistent.
For $\delta > \varrho$, a better approximation, independent of the chosen kernel function, can be derived from the analytical solution for $r > \varrho$, \cref{eq:ana_constkernel}, yielding
\begin{equation}
\label{eq:corrected_xi_larger}
\Xi_\delta = \left\{ 1 + \frac{R L_p\mu_I}{\kappa}\ln\left(\frac{\delta}{R}\right) \right\}^{-1} \quad \text{for}\; \delta > \varrho.
\end{equation}
Note that this $\Xi_\delta$ is a sensible flux correction for the methods $\textsc{ls}$ and $\textsc{css}$, too. However, herein,
we do not modify $\hat{q}_m$ for these methods and retain the methods as described in \citep{d2007multiscale} and \citep{koeppl2018}.

\subsection{Kernel integration}

The kernel integral in \cref{eq:qkomegakernel}
\begin{equation}
\mathbb{I}_{\Phi,K_\Omega} := \int_{K_\Omega \cap \mathcal{S}(\Phi_\Gamma)} \Phi_\Gamma \,\text{d}x,
\end{equation}
is hard to approximate with standard quadrature rules, since $K_\Omega \cap \mathcal{S}(\Phi_\Gamma)$ is difficult to compute. However, the integral over the entire support $\mathcal{S}(\Phi_\Gamma)$ is known exactly. Hence, the problem can be reformulated as distributing the whole integral over the control volumes $K_\Omega$ weighted with the respective support volume fractions. To this end, we create $n_\Gamma$ points $\boldsymbol{x}_i \in \mathcal{S}(\Phi_\Gamma)$ with associated volume elements $V_i$ of similar size and shape, so that
\begin{equation}
\mathbb{I}_{\Phi,K_\Omega} \approx \sum\limits_{i=1, \boldsymbol{x}_i\in K_\Omega}^{n_\Gamma} V_i\Phi_\Gamma(\boldsymbol{x}_i), \quad \sum\limits_{i=1}^{n_\Gamma} V_i = \vert\mathcal{S}(\Phi_\Gamma)\vert.
\end{equation}

\section{Numerical experiments}
\label{sec:numerical}

The three considered methods, $\textsc{ls}$, $\textsc{css}$, $\textsc{ds}$,
yield different solutions for $p_t$ and $r<R$ (\textsc{ls}, \textsc{css}), or $r<\varrho$ (\textsc{ds}).
However, the solutions for $p_v$ and $\hat{q}_m$ are identical. We denote the different pressure
solutions in $\Omega$ by $p_t^\mathbb{M}$, $\mathbb{M} \in \{ \textsc{ls}, \textsc{css}, \textsc{ds}\}$.
We analyze these methods with different vessel configurations in a series of numerical experiments.
In \cref{sec:numerical_single}, we consider a single straight vessel.
The numerical methods are investigated in terms of the ratio of grid resolution to vessel radius, comparing with analytical solutions.
In \cref{sec:numerical_multiple}, we construct an analytical solution for three parallel vessels, that show
that the optimal flux scaling factor $\Xi$ is independent of perturbations caused by neighboring vessels.
For each numerical experiment, the setup is described and the results are presented and discussed.

\subsection{Single vessel}
\label{sec:numerical_single}

\begin{figure}
\centering
\includegraphics[width=\textwidth]{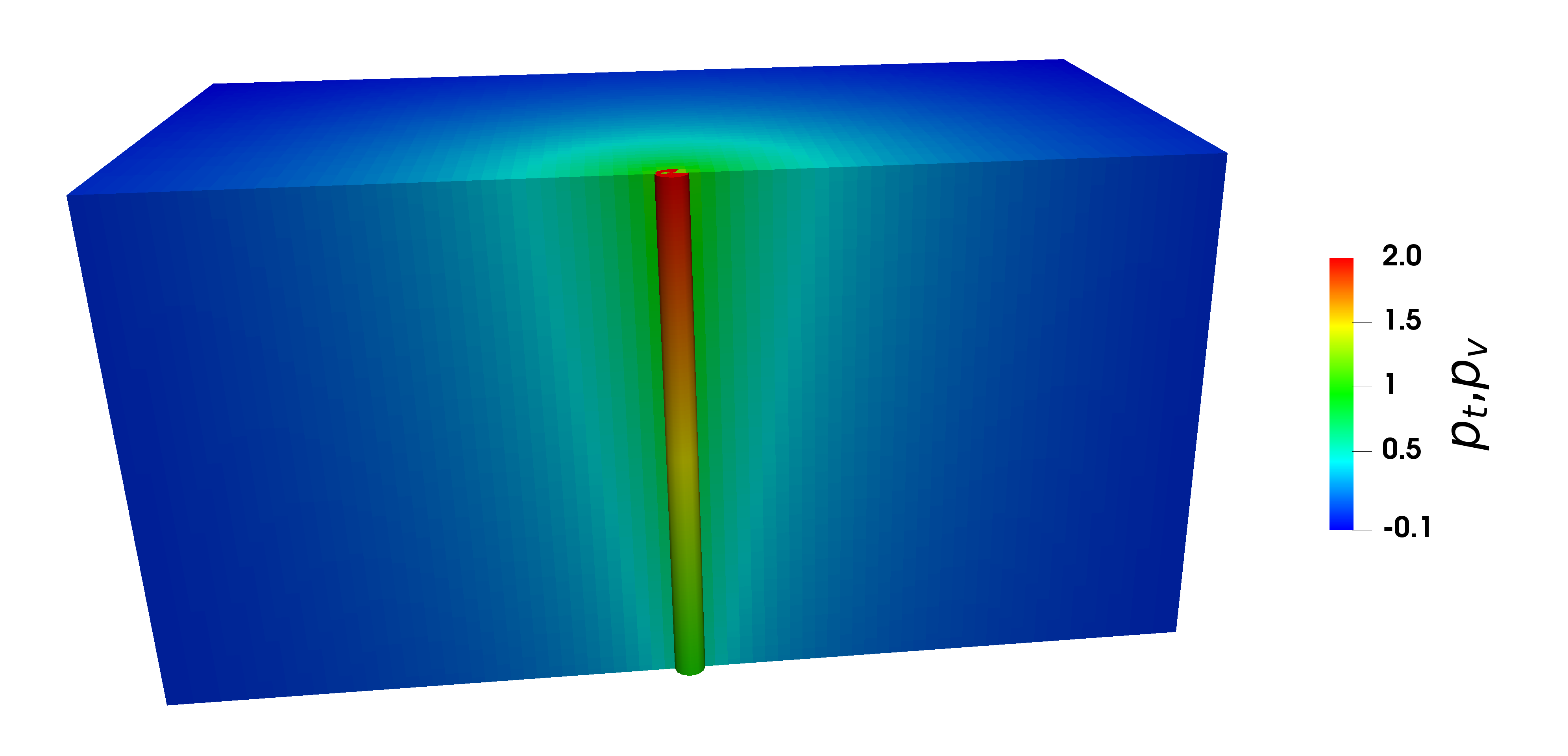}
\caption{Numerical solution $p^{\textsc{ds}}_t$ for $R = 0.03$ and $\varrho = R$. The grid resolution in the extra-vascular domain is $80\times80\times80$,
so that $h = 0.025$. The vascular pressure solution $p_v$ is projected on a tube with radius $R$.}
\label{fig:ex1_paraview}
\end{figure}

Let us consider a slightly simplified problem, adapted from \cite{d2007multiscale},
\begin{subequations}
\label{eq:problembeta}
\begin{align}
  -\frac{\partial}{\partial s} \left(\kappa_v \frac{\partial p_v}{\partial s} \right) &= -\hat{q}_m \quad&& \text{in } \Lambda,\\
  -\laplace p_t &= \hat{q}_m\Phi_\Lambda \quad&& \text{in } \Omega, \\
  \hat{q}_m &= \beta (p_v - p_{t,0})\Xi,
\end{align}
\end{subequations}
with the domains $\Omega = [0,1]\times[0,1]\times[0,1]$ and $\Lambda = \{0.5\}\times\{0.5\}\times[0,1]$,
i.e the vessel center-line coincides with the $z$ axis.
By choosing the parameters as
\begin{equation*}
\kappa_v = 1 + z + \frac{1}{2}z^2, \quad \beta = \frac{2\pi}{2\pi + \ln R},
\end{equation*}
the pressure solutions,
\begin{subequations}
\label{eq:problembeta_sol}
\begin{align}
p_{v,e} &= 1 + z, \\
p_{t,e}^{\textsc{ls}} &=  - \frac{1 + z}{2\pi} \ln r, \\
p_{t,e}^{\textsc{css}} &= \begin{cases} - \frac{1 + z}{2\pi} \ln R & r \leq R, \\ - \frac{1 + z}{2\pi} \ln r & r > R, \end{cases} \\
p_{t,e}^{\textsc{ds}} &= \begin{cases} - \frac{1 + z}{2\pi} \left[ \frac{r^2}{2\varrho^2} + \ln\left(\frac{\varrho}{R}\right) - \frac{1}{2} \right] & r \leq \varrho, \\ - \frac{1 + z}{2\pi} \ln r & r > \varrho, \end{cases}
\end{align}
\end{subequations}
with $r = \sqrt{x^2 + y^2}$, solve \cref{eq:problembeta} given the boundary conditions
\begin{align*}
  p_v &= 1 \quad&& \text{on } \partial\Lambda \cap \{z=0\},\\
  p_v &= 2 \quad&& \text{on } \partial\Lambda \cap \{z=1\},\\
  \grad p_t \cdot \boldsymbol{n} &= - \frac{1}{2\pi} \ln r \quad&& \text{on } \partial\Omega \cap \{z=1\}, \numberthis \label{eq:problembeta_bc}\\
  \grad p_t \cdot \boldsymbol{n} &= \frac{1}{2\pi} \ln r \quad&& \text{on } \partial\Omega \cap \{z=0\}, \\
  p_t &= - \frac{1 + z}{2\pi} \ln r \quad&& \text{on } \partial\Omega \setminus \{z=0, z=1\},
\end{align*}
where $\boldsymbol{n}$ is the outward-pointing unit normal vector on the boundary $\partial\Omega$ of the domain $\Omega$. From the analytical
pressure solutions follows that $\hat{q}_{m,e} = 1+z$ is the analytical source term.

The pressure discretization errors are computed in the normalized discrete norm
\begin{equation}
 \vert\vert p_t - p_{t,e} \vert\vert_2 := \frac{\sum_{\Omega_h} \vert K_\Omega \vert (p_{K_\Omega,e} - p_{K_\Omega})^2}{\sum_{\Omega_h} \vert K_\Omega \vert},
\end{equation}
where $p_{K_\Omega}$, $p_{K_\Omega,e}$ denote numerical and exact pressure evaluated at the center of a control volume
$K_\Omega$ and $\vert K_\Omega \vert$ its volume. The error for $p_v$ in $\Lambda_h$ is computed analogously.
The error in the source term $\hat{q}_{m}$ is computed as
\begin{equation}
 \vert\vert q - q_{e} \vert\vert_2 = \frac{\sum_{\Lambda_h} \vert K_\Lambda \vert (q_{K_\Lambda,e} - q_{K_\Lambda})^2}{\sum_{\Lambda_h} \vert K_\Lambda \vert},
\end{equation}
where
\begin{equation}
 q_{K_\Lambda,e} = \int_{K_\Lambda} \!\hat{q}_{m,e}\,\text{d}s \quad \text{and} \quad q_{K_\Lambda} = \int_{K_\Lambda}\! \hat{q}_m \,\text{d}s.
\end{equation}
The maximum control volume size, $h$, is given by the maximum edge length in both domains.
We choose the edge length so that $h = h_\Omega$ = $h_\Lambda$.

\begin{figure}
\centering
\includegraphics[width=\textwidth]{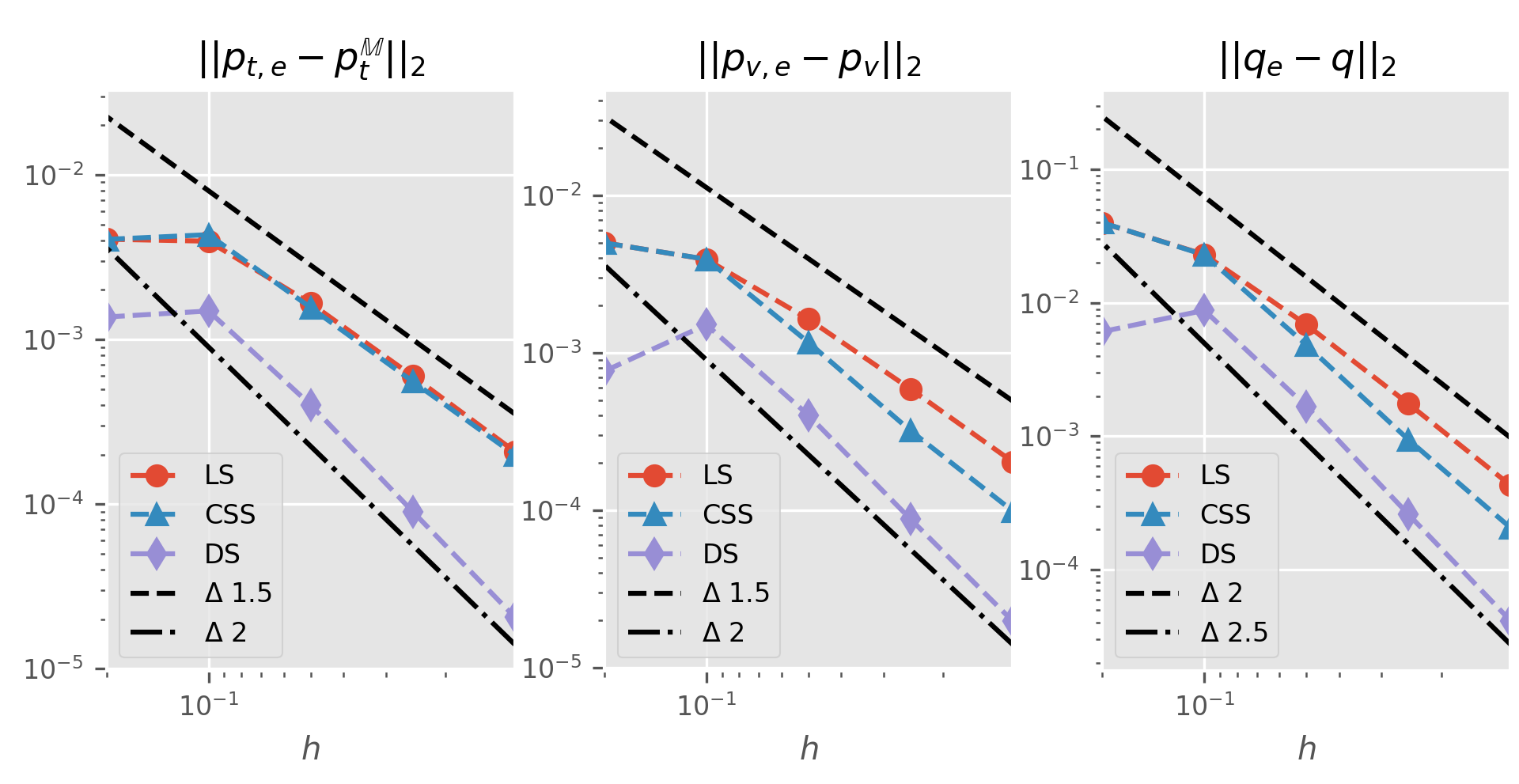}
\caption{Discretization errors in $p_t$, $p_v$, and $\hat{q}_m$ for the different methods \textsc{ls}, \textsc{css}, \textsc{ds}, with vessel radius $R=0.1$,
and kernel support radius $\varrho=R$ (only \textsc{ds}). The exact solution $p_{t,e}$ is the analytical solution $p^{\textsc{ls}}_t(\boldsymbol{x})$, $p^{\textsc{css}}_t(\boldsymbol{x})$, or $p^{\textsc{ds}}_t(\boldsymbol{x})$, corresponding to the respective method.
The black lines are curves with slopes of $1.5$, $2$, and $2.5$ for comparison.}
\label{fig:rates_r01}
\end{figure}

\begin{figure}
\centering
\includegraphics[width=\textwidth]{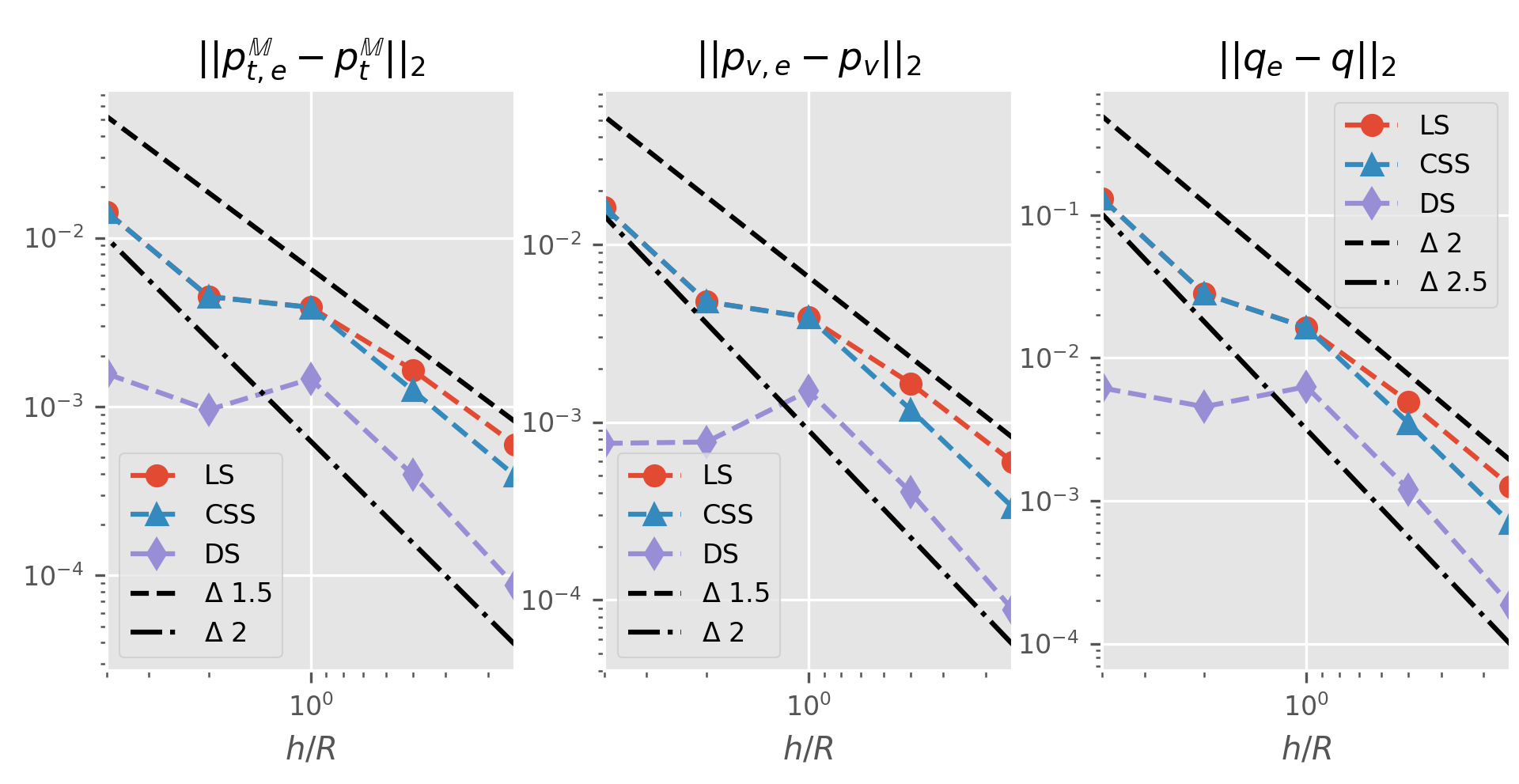}
\caption{Discretization errors in $p_t$, $p_v$, and $\hat{q}_m$ for the different methods \textsc{ls}, \textsc{css}, \textsc{ds}, with vessel radius $R=0.05$,
and kernel support radius $\varrho=R$ (only \textsc{ds}). The exact solution $p_{t,e}$ is the analytical solution $p^{\textsc{ls}}_t(\boldsymbol{x})$, $p^{\textsc{css}}_t(\boldsymbol{x})$, or $p^{\textsc{ds}}_t(\boldsymbol{x})$, corresponding to the respective method.}
\label{fig:rates_r005}
\end{figure}

The numerical solutions $p^{\textsc{ds}}_t$, $p_v$, for $R = 0.03$ and $\varrho = R$, are exemplarily shown in \cref{fig:ex1_paraview}.
The discretization error and the convergence rates are computed for $p_t$, $p_v$, and $q$.
\cref{fig:rates_r01} shows the discretization errors for $R=0.1$, $\varrho=R$ and $\Phi^\text{const}$. It can be seen that $\textsc{ds}$ is the
only method with optimal convergence rate for $p_t$ in the given norm. The error in $p_t$ cannot directly be compared, since it is
computed with respect to the respective analytical solution corresponding to the chosen method, which differ for $r < \varrho$. However, all
methods are expected to converge to the same analytical solution for $p_v$ and $q$. It is evident from the error plots of $p_v$ and $q$ that
$\textsc{ds}$ shows the lowest discretization error of all three methods. Furthermore, for the presented numerical experiment $\textsc{ds}$ achieves
convergence rates in $q$ of approx. $2.5$, while $\textsc{ls}, \textsc{css}$ show convergence rates of approx. $2$. We conclude that the increased
discretization error for $\textsc{ls}$ and $\textsc{css}$ in $p_t$, influences the approximation of $p_v$, $q$, and is likely due to the insufficient
approximation of $p_{t,\mathbb{W}}$.
\cref{fig:rates_r005} shows the discretization errors for $R=0.05$, $\varrho=R$ and $\Phi^\text{const}$. The first two error measurements show the situation $h > \varrho$.
If the discretization length $h$ is larger than the kernel radius $\varrho$, the three methods do not differ in the representation of the source term. However, for
the $\textsc{ds}$ method, we introduced a source term correction by an adjusted flux scaling factor dependent on the discrete distance $\delta$ (see~\cref{eq:corrected_xi}), which is proportional to $h$. It can be seen, that this adjustment significantly reduces the error for all quantities. Note that this scaling factor for $r>R$, can also be
applied for the other methods ($\textsc{ls}$ and $\textsc{css}$), however its motivation is directly drawn from the new formulation of the perfusion problem in \cref{eq:kernel}.

\FloatBarrier
\subsection{Multiple parallel vessels}
\label{sec:numerical_multiple}

\begin{figure}
\centering
\includegraphics[width=\textwidth]{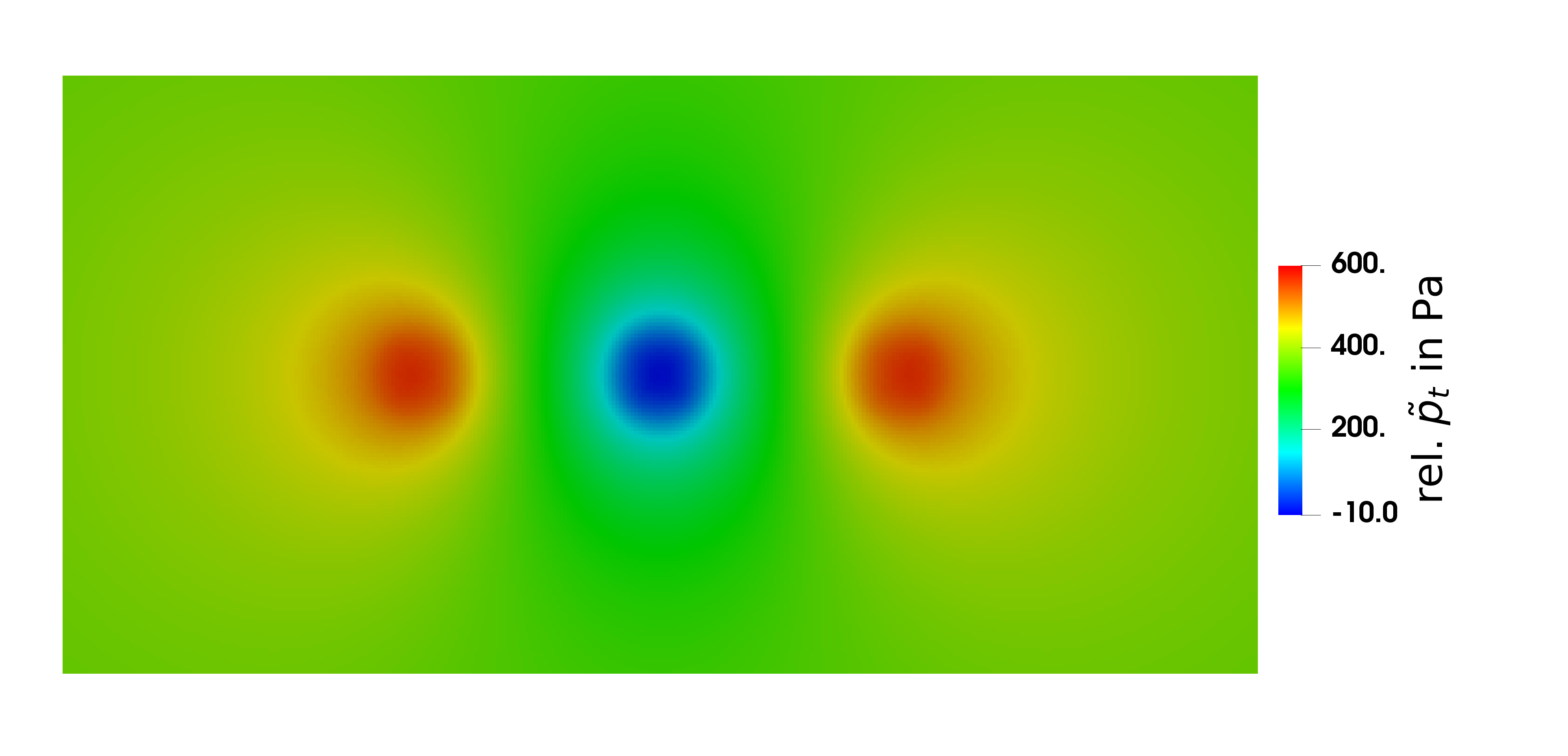}
\caption{Numerical solution of hydraulic pressure $\tilde{p}^{\textsc{ds}}_t$ (relative to $\tilde{p}_\text{atm} = \SI{1e5}{\Pa}$) for a kernel support radius $\varrho = 2R$, where $R = R_1 = R_2 = R_3 = \SI{4}{\micro\m}$.
The grid resolution in the extra-vascular domain is $320\times160$, so that $h = \SI{0.625e-6}{\micro\m}$.}
\label{fig:ex2_paraview}
\end{figure}

In the next experiment, we consider three parallel vessels, from which two are emitting fluid (arterial side), and one vessel is absorbing fluid (venous side).
We assume constant pressures in the vessels, $\tilde{p}_{v,1} = \tilde{p}_{v,3} = \SI{3400}{\Pa}$, $\tilde{p}_{v,2} = \SI{2300}{\Pa}$, which renders the problem effectively two-dimensional. Then, exploiting the linearity of the Laplace operator, we can construct a solution of \cref{eq:linesources} in $\Omega$
using the superposition principle, i.e.,
\begin{equation}
p^{\textsc{ls}}_t(\boldsymbol{x}) = \sum\limits_{i=1}^3 p_{t,i} = \sum\limits_{i=1}^3 -\frac{\mu_I \hat{q}_{m,i}}{2\pi \kappa \rho_I} \ln r_i, \quad r_i = \vert\vert \boldsymbol{x}_i - \boldsymbol{x} \vert\vert_2,
\end{equation}
where $\boldsymbol{x}_i$ is the position of the center-line of vessel $i$, and $\boldsymbol{x} = [x,y,z]^T$ a point in $\Omega$.
Recall that $\hat{q}_{m,i} = \rho_I L_p 2 \pi R_i (p_{v,i} - p_{t,\mathbb{W}_i})$ is a linear function in the arguments $p_{v,i}$ and $p_{t,\mathbb{W}_i}$, and
\begin{equation}
p_{t,\mathbb{W}_i} = \frac{1}{2\pi}\int_0^{2\pi} \left. p^{\textsc{ls}}_t \right\rvert_{R_i} \text{d}\theta,
\end{equation}
is the mean pressure on the surface of vessel $i$. Thus, using the mean value property of harmonic functions yields
\begin{equation}
\label{eq:exact_multiple}
p_{t,\mathbb{W}_i} = \sum\limits_{\substack{j=1 \\ j \neq i}}^3 \left( -\frac{\mu_I \hat{q}_{m,j}}{2\pi \kappa \rho_I} \ln r_{ij} \right)  -\frac{\mu_i \hat{q}_{m,i}}{2\pi \kappa \rho_I} \ln R_i,
\end{equation}
where $r_{ij}$ denotes the Euclidean distance of the center-lines of vessels $i$ and $j$ and $R_i$ the radius of vessel $i$.
\cref{eq:exact_multiple} constitutes a system of three linear equations with the unknowns $p_{t,\mathbb{W}_1}$, $p_{t,\mathbb{W}_2}$, $p_{t,\mathbb{W}_3}$.
As the analytical solution of \cref{eq:exact_multiple} results in a rather lengthy expression, we compute $p_{t,\mathbb{W}_i}$ numerically. Using the parameter values, $r_{12}=r_{23}=\SI{40}{\micro\m}$, $r_{12}=\SI{80}{\micro\m}$, $R_1 = R_2 = R_3 = \SI{4}{\micro\m}$,
$L_p = \SI{1e-9}{\m\per\pascal\s}$, $\kappa = \SI{8.3e-18}{\m\squared}$, $\mu_I = \SI{1.339e-3}{\pascal\s}$,
$\rho_I = \SI{1030}{\kg\per\cubic\m}$, $\Pi_t = \SI{666}{\pascal}$, $\Pi_v = \SI{3300}{\pascal}$, $\sigma = 1.0$, cf.~\citep{Koch2018b}, we obtain
$\tilde{p}_{t,\mathbb{W}_1} = \tilde{p}_{t,\mathbb{W}_3} \approx \SI{587.29}{\pascal}$, and $\tilde{p}_{t,\mathbb{W}_2} \approx \SI{-89.350}{\pascal}$. Recall that
$p_{t/v} = \tilde{p}_{t/v} - \Pi_{t/v}$.

\begin{figure}
\centering
\includegraphics[width=\textwidth]{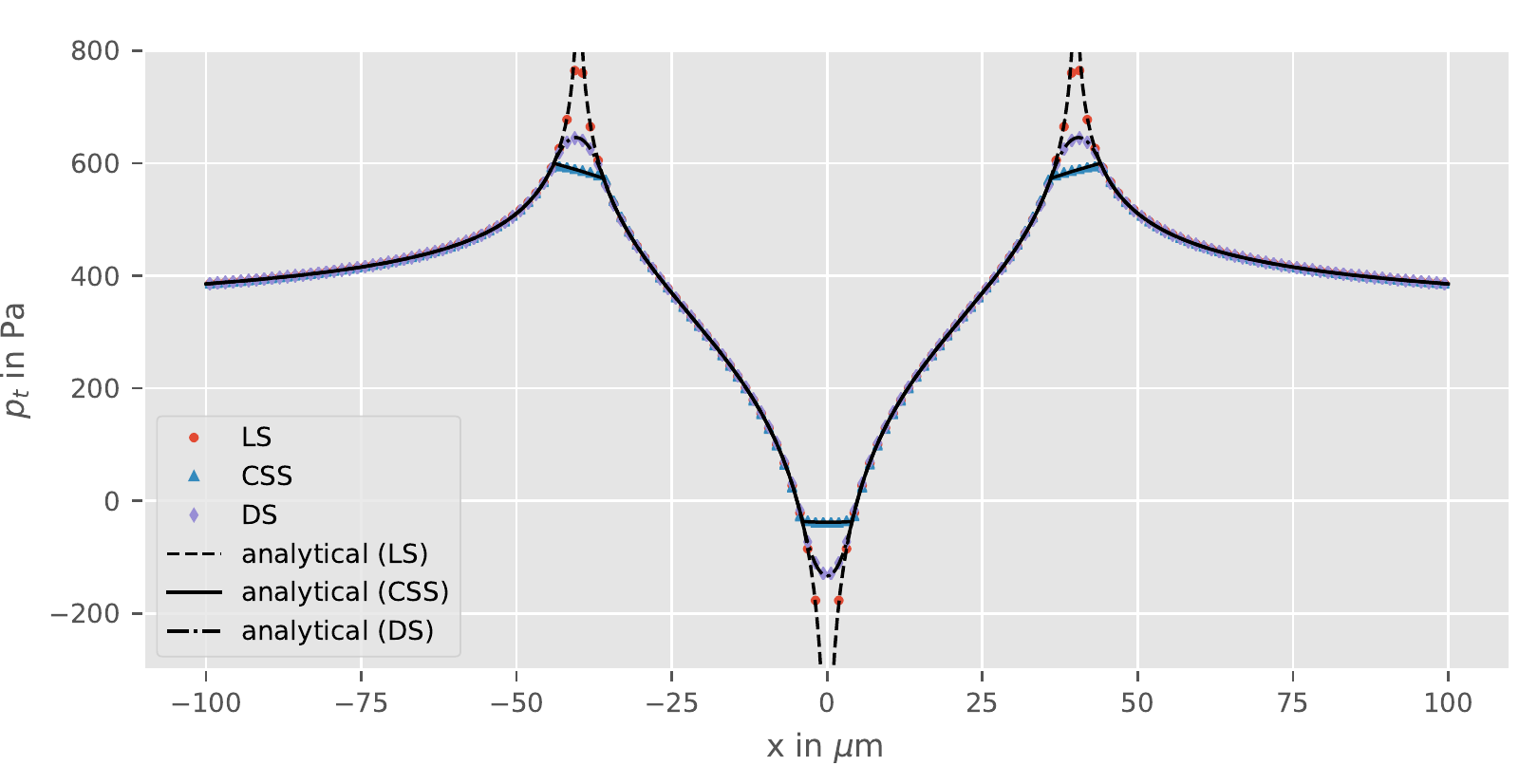}
\caption{Analytical solutions (black lines) $p^{\textsc{ls}}_t(\boldsymbol{x})$, $p^{\textsc{css}}_t(\boldsymbol{x})$, and $p^{\textsc{ds}}_t(\boldsymbol{x})$,
         and the corresponding numerical solutions (colored markers) obtained with a grid resolution of $160\times 80$ for the three methods.
         The solution for $\textsc{ds}$ use a kernel support radius of $\varrho=R$. The middle peak corresponds to the fluid-absorbing vessel,
         whereas the left and the right peak correspond to the fluid-emitting vessels.}
\label{fig:exact_multiple}
\end{figure}

\begin{figure}
\centering
\includegraphics[width=\textwidth]{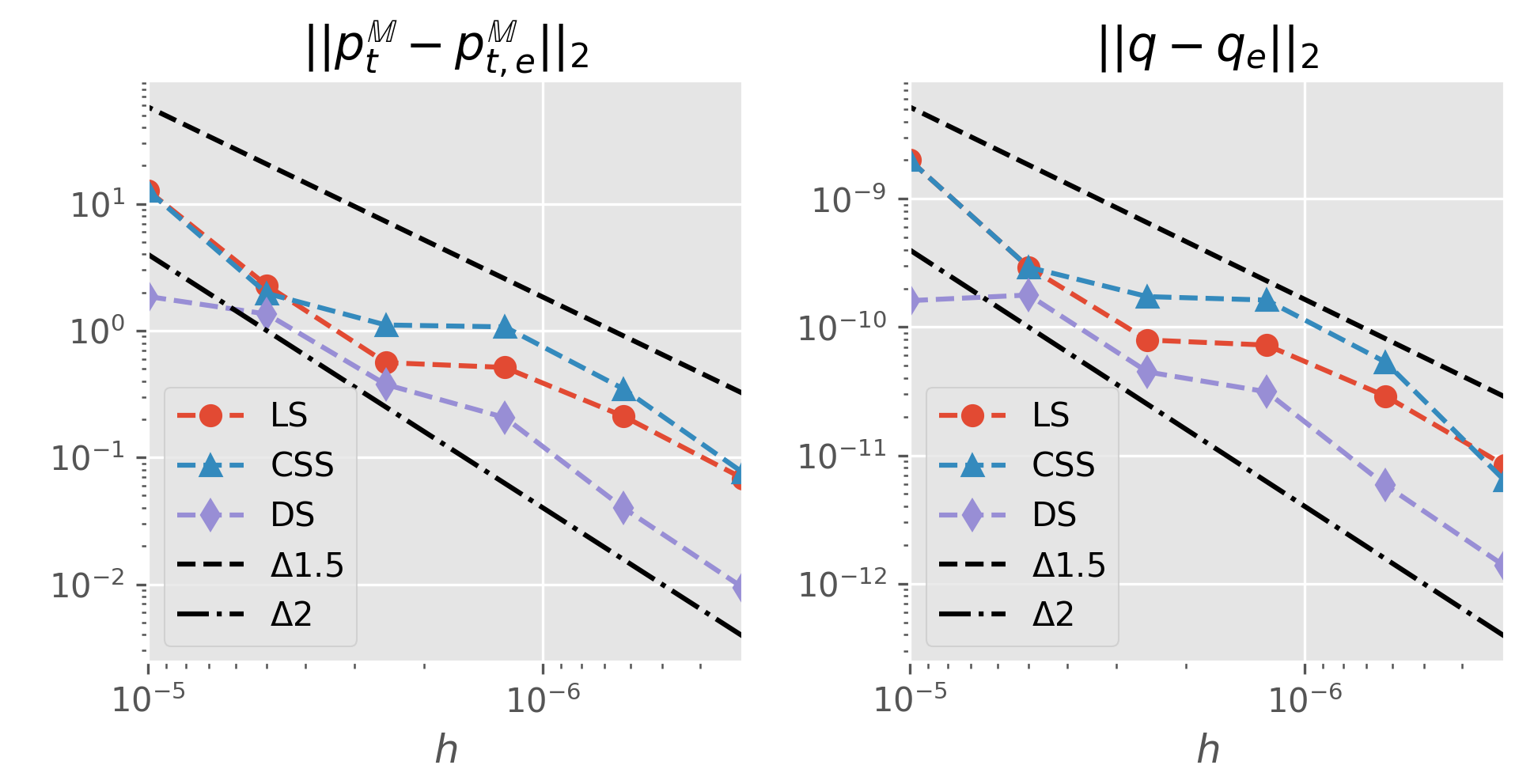}
\caption{Discretization errors in $p_t$ and $\hat{q}_m$ for the different methods \textsc{ls}, \textsc{css}, and \textsc{ds} with kernel support radius $\varrho=R$ (only \textsc{ds}).}
\label{fig:multivessel_rates}
\end{figure}

Again, the solution
with line sources is equal to the solution with distributed sources outside the kernel support radius. Furthermore, the superposition principle
equally applies for the distributed source model so that, for instance, for the kernel function $\Phi^\text{const}$,
\begin{equation}
p^{\textsc{ds}}_t(\boldsymbol{x}) = \begin{cases}
    \sum\limits_{i=1}^3 -\frac{\mu_I \hat{q}_{m,i}}{2\pi \kappa \rho_I} \ln r_i & r_i > \varrho_i,\\
    \sum\limits_{\substack{j=1 \\ j \neq i}}^3 -\frac{\mu_I \hat{q}_{m,j}}{2\pi \kappa \rho_I} \ln r_j -\frac{\mu_I \hat{q}_{m,i}}{2\pi \kappa \rho_I}
    \left[ \frac{r_i^2}{2\varrho_i^2} + \ln \varrho_i - \frac{1}{2} \right] & r_i \leq \varrho_i,
    \end{cases}
\end{equation}
is a solution to \cref{eq:kernel}, given that the kernel support regions of the vessels do not overlap.
For the \textsc{css} method the solution is given
by a linear continuation of the pressure for $r_i \leq R_i$, cf. \citep{koeppl2018},
\begin{equation}
p^{\textsc{css}}_t(\boldsymbol{x}) = \begin{cases}
    \sum\limits_{i=1}^3 -\frac{\mu_I \hat{q}_{m,i}}{2\pi \kappa \rho_I} \ln r_i & r_i > R_i,\\
    \sum\limits_{\substack{j=1 \\ j \neq i}}^3 -\frac{\mu_I \hat{q}_{m,j}}{2\pi \kappa \rho_I} \ln r_j -\frac{\mu_I \hat{q}_{m,i}}{2\pi \kappa \rho_I}
    \ln R_i & r_i \leq R_i.
    \end{cases}
\end{equation}
The analytical solutions $p^{\textsc{ls}}_t(\boldsymbol{x})$, $p^{\textsc{css}}_t(\boldsymbol{x})$, $p^{\textsc{ds}}_t(\boldsymbol{x})$
for a domain $\Omega = [-100,100]\times[-50,50]\si{\micro\m}$ with $\boldsymbol{x}_1 = [-40, 0]^T \si{\micro\m}$,
$\boldsymbol{x}_2 = [0, 0]^T \si{\micro\m}$, $\boldsymbol{x}_3 = [40, 0]^T \si{\micro\m}$ along the x-axis are shown
in \cref{fig:exact_multiple}. The numerical solution $\tilde{p}^{\textsc{ds}}_t = p^{\textsc{ds}}_t - \pi_t$,
for $\varrho = 2R$, is exemplarily shown in \cref{fig:ex2_paraview}.
The discretization errors with respect to the analytical solutions for $p^\mathbb{M}_t$ and $q$ are shown in \cref{fig:multivessel_rates}.
As for the numerical experiment with a single vessel, method $\textsc{ds}$ shows the lowest error for the multi-vessel experiment.
Additionally, it can be seen that the flux-scaling for $r>\varrho$ significantly improves the approximation of $q$, although
the analytical solution is no longer strictly radial around the individual vessels.

\begin{figure}
\centering
\includegraphics[width=\textwidth]{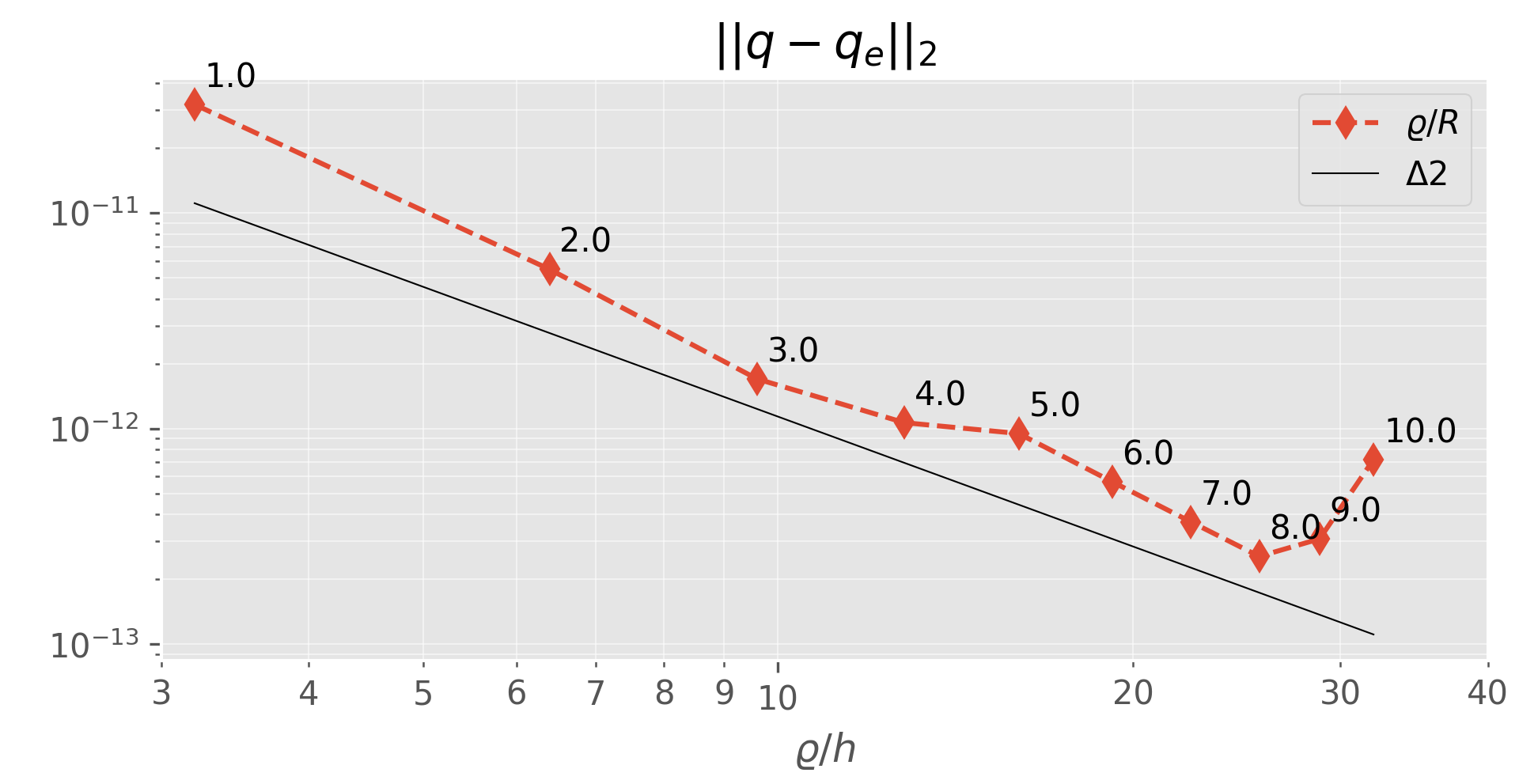}
\caption{Discretization errors in $\hat{q}_m$ for different kernel support radii $\varrho$. The annotated values show the ratio of the kernel support radii to the vessel radii $\varrho/R$. The black line has a slope of $2$ and is shown for reference.}
\label{fig:multivessel_rho}
\end{figure}

In a second experiment, the grid resolution is fixed to $h = \SI{1.25}{\micro\m}$.
Then, the kernel support radius is step-wise increased starting from $\varrho=R$,
for all vessels. The discretization errors for $q$ are shown in \cref{fig:multivessel_rho}.
Recall that the distances $r_{12} = r_{23} = \SI{40}{\micro\m}$, such that the kernel support region for two neighboring vessels start intersecting
for $\varrho > 5R$, and the kernel support region includes the location of the center-line of the neighboring vessel for $\varrho > 10R$.
It can be seen that with increasing kernel radius the approximation of the source term improves significantly.
Comparing \cref{fig:multivessel_rho} and \cref{fig:multivessel_rates} it seems that an increase of the kernel support radius has the same effect as increasing
the grid resolution. An increase of the kernel support area by a factor of $2$ reduces the discretization error by a factor of $2$.
However, note that the pressure solution is increasingly regularized and thus deviates from the physically sensible solution.
If the kernel support radius is chosen too large, the regularization affects neighboring vessels, such that the error in the source term increases again.
The results suggest that the variable kernel support region, to some extent, decouples the source term approximation error from the 3D grid resolution.

\FloatBarrier
\subsection{Multiple arbitrarily-oriented vessel}
\label{sec:numerical_multiple3d}

\begin{figure}
\centering
\includegraphics[width=0.8\textwidth]{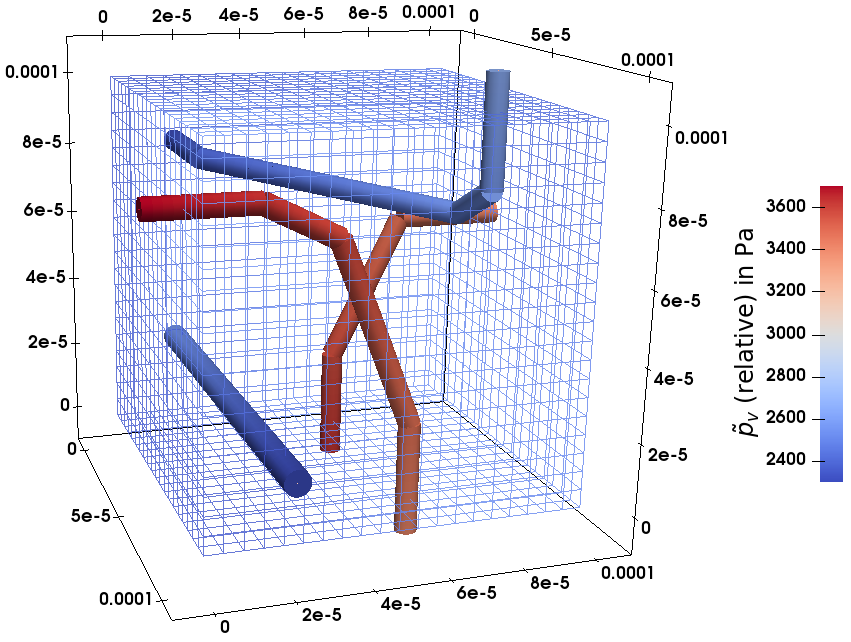}
\caption{Computational domain for a numerical experiment with multiple arbitrarily-oriented vessels.
         The vessel segments are shown as cylinders scaled with the respective vessel radius. The vessel segment color corresponds to the relative hydraulic vessel pressure $\tilde{p}_v$. The cubic extra-vascular domain has the dimensions $\SI{100x100x100}{\micro\meter}$.}
\label{fig:multivessel3d}
\end{figure}

In this experiment, we consider multiple arbitrarily-oriented vessels embedded in a cubic extra-vascular domain, for which no analytical solution is given.
Recall that for such a system, the methods \textsc{ls}, \textsc{css} and \textsc{ds} do not generally give the same solutions, but the differences are expected
to be small.
In order to isolate differences stemming from the different source models from other sources of error, the vessels do not bifurcate and are constructed in a way that they intersect the 3D domain boundary perpendicularly.
The computational domain is shown in~\cref{fig:multivessel3d}.
Two arterial (fluid-emitting) and two venous (fluid absorbing) capillaries are embedded in a cubic extra-vascular domain with dimensions $\SI{100x100x100}{\micro\meter}$.
For each vessel, one end is chosen as inflow and the other as outflow boundaries. On the inflow boundary of vessel $i$, a constant inflow rate, $q_{v,i,\text{in}} = \rho_B \pi R_i^2 v_i$ is enforced, where $R_i$ is the inflow segment radius and $v_i$ the inflow velocity.
On outflow boundaries, we fix the pressure, $\tilde{p}_{v,i,\text{out}}$. The data for the geometry of the vessel and boundary conditions is given in \cref{tab:vesseldata} in the appendix. The remaining model parameters are chosen as in the previous multiple vessel experiment.
The boundaries of the extra-vascular domain are considered symmetry boundaries, hence $\frac{\partial p_t}{\partial\boldsymbol{n}} = 0$ on $\partial\Omega$.

\begin{figure}
\centering
\includegraphics[width=\textwidth]{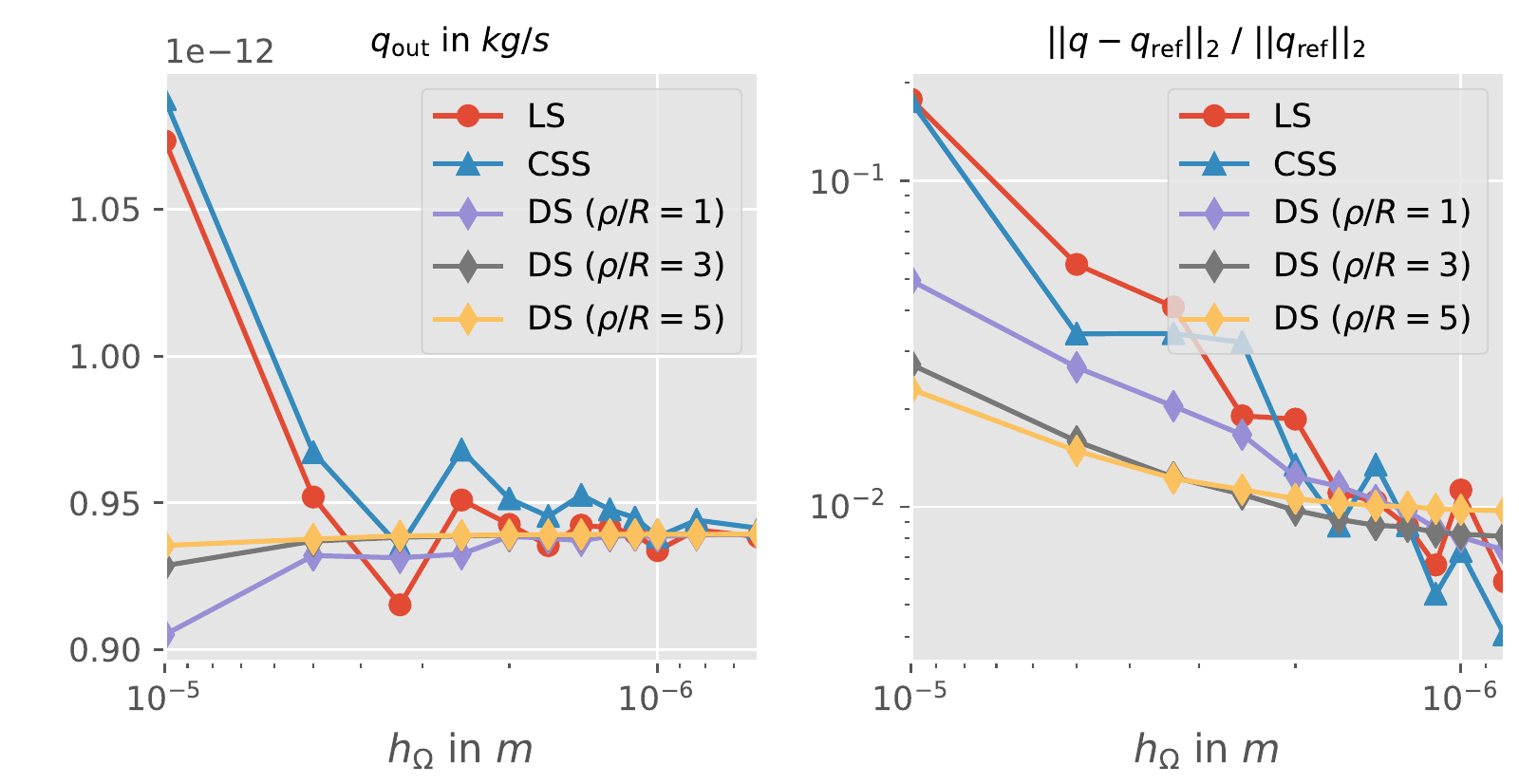}
\caption{Left, the total mass flux emitted by the arterial vessel $q_\text{out}$ for the different methods (\textsc{css}, \textsc{ls}, and \textsc{ds}) with grid refinement. Right, the difference in the source term between the different methods with grid refinement with respect to the reference solution (\textsc{css}).}
\label{fig:multivessel3d_results}
\end{figure}

We produce a reference solution using the \textsc{css} method, with $h_\Omega = \SI{0.625}{\micro\m}$, $h_\Lambda = \SI{0.5}{\micro\m}$. The source term
$\hat{q}_m$ is computed for every cell $K \in \Lambda_h$, resulting in a source vector $q_\text{ref}$.
Then, the corresponding source term, $q$, for $h_\Lambda = \SI{0.5}{\micro\m}$, is computed for different $h_\Omega$ using the methods \textsc{css}, \textsc{ls}, and \textsc{ds} with different kernel radii $\varrho/R_i = 1, 3$, and $5$. Furthermore, we compute the total mass flux, $q_\text{out}$, emitted by the arterial vessels, as the sum of all fluxes leaving the vessel domain into the extra-vascular domain.
The results are shown in~\cref{fig:multivessel3d_results}.
First, it can be seen that with a coarse grid, the difference in the source term computed by the different schemes is quite large ($\approx \SI{10}{\percent}$ relative to the reference solution). Notably, the lowest difference with respect to the reference solution at coarse resolutions is achieved by the \textsc{ds} with the largest kernel support region.
With grid refinement these differences decrease to less than \SI{1}{\percent}.
However, it can be seen that the curve flattens for the \textsc{ds} methods with larger kernel for fine grid resolutions.
The results suggest that the difference cannot be reduced to much less than \SI{1}{\percent}.
At the given grid resolutions, such a behavior is not observed yet for the \textsc{ds} with the smallest kernel support.

We conclude that the flattening of the error curve for the \textsc{ds} methods with larger kernel is rather caused by overlapping
kernels of neighboring vessels as well as at bends. This effect, which is further investigated in the next numerical example, is minimized for the smallest kernel support.
From the fact, that the kernel curve still suggests convergence until a very fine grid resolution leads us to the conclusion that the error caused by the approximation of the
flux scaling factor $\Xi$ (as discussed in \cref{sec:theory_multiple}) is much smaller.
In perspective of the rather big uncertainties stemming from vessel segmentation and modeling error,
the difference of \SI{1}{\percent} between the different methods present in \cref{fig:multivessel3d_results} will most certainly be acceptable in practical simulations.
Furthermore, the observations in this experiment support our results from the previous experiment that for coarser mesh resolutions
it is better to choose a larger kernel support, if a good approximation of the source term $\hat{q}_m$ is important.

\FloatBarrier
\subsection{Vessel network}
\label{sec:numerical_network}

\begin{figure}
\centering
\includegraphics[width=\textwidth]{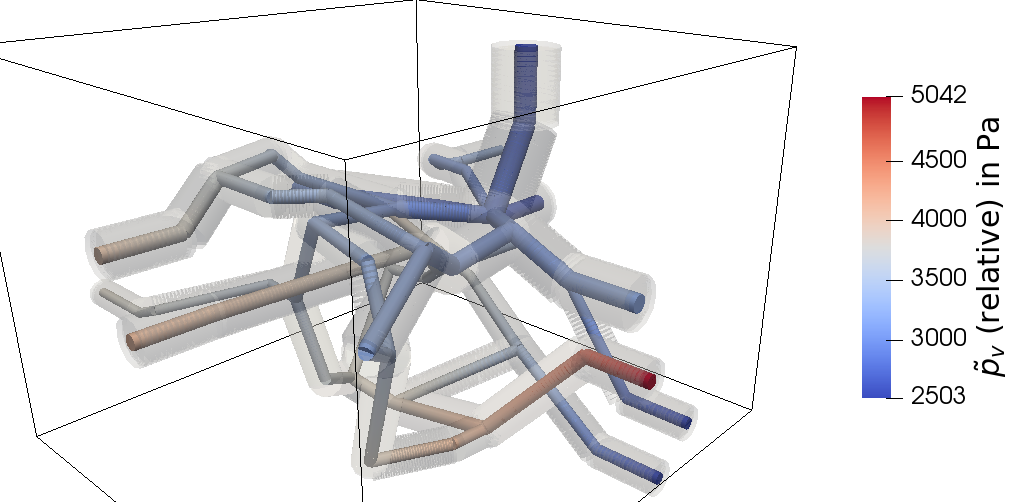}
\caption{Vessel network from the superficial cortex of a rat brain~\citep{motti1986,secomb2000theoretical}. The vessel segments are shown as cylinders scaled with the respective vessel radius. The vessel segment color corresponds to the relative hydraulic vessel pressure $\tilde{p}_v$. The cubic extra-vascular domain has the dimensions $\SI{200x210x190}{\micro\meter}$.
The kernel support volume for $\varrho_i = 3R_i$ is visualized in opaque gray.}
\label{fig:network}
\end{figure}
In the last numerical experiment, we consider a network of capillaries extracted from the superficial cortex of a rat brain~\citep{motti1986,secomb2000theoretical}.
Inlets and outlets are annotated in the data set.
For the inlets, velocity estimates based on the vessel radius are given in~\citep{secomb2000theoretical}, and herein enforced as Neumann boundary conditions.
The vessel radii are in the range of \SIrange{2}{4.5}{\micro\m}.
Dirichlet boundary conditions enforce $p_{v,\text{out}} = \SI{1.025e5}{\pascal}$ at the outlets. The extra-vascular domain $\Omega$ is given by
a rectangular box, $\SI{200}{\um} \times \SI{210}{\um} \times \SI{190}{\um}$. All boundaries $\partial\Omega$ are considered symmetry boundaries,
$\frac{\partial p_t}{\partial\boldsymbol{n}} = 0$ on $\partial\Omega$. The network boundaries are extended by \SI{30}{\micro\m} segments
with perpendicular intersections on $\partial\Omega$. This adjustment to the network structure
is necessary to better match the assumption of symmetry boundaries in the extra-vascular domain. Vessels intersecting the boundary at acute angles
lead to non-physical, very non-radial flows around the vessel end at symmetry boundaries on $\partial\Omega$.

\begin{figure}
\centering
\includegraphics[width=\textwidth]{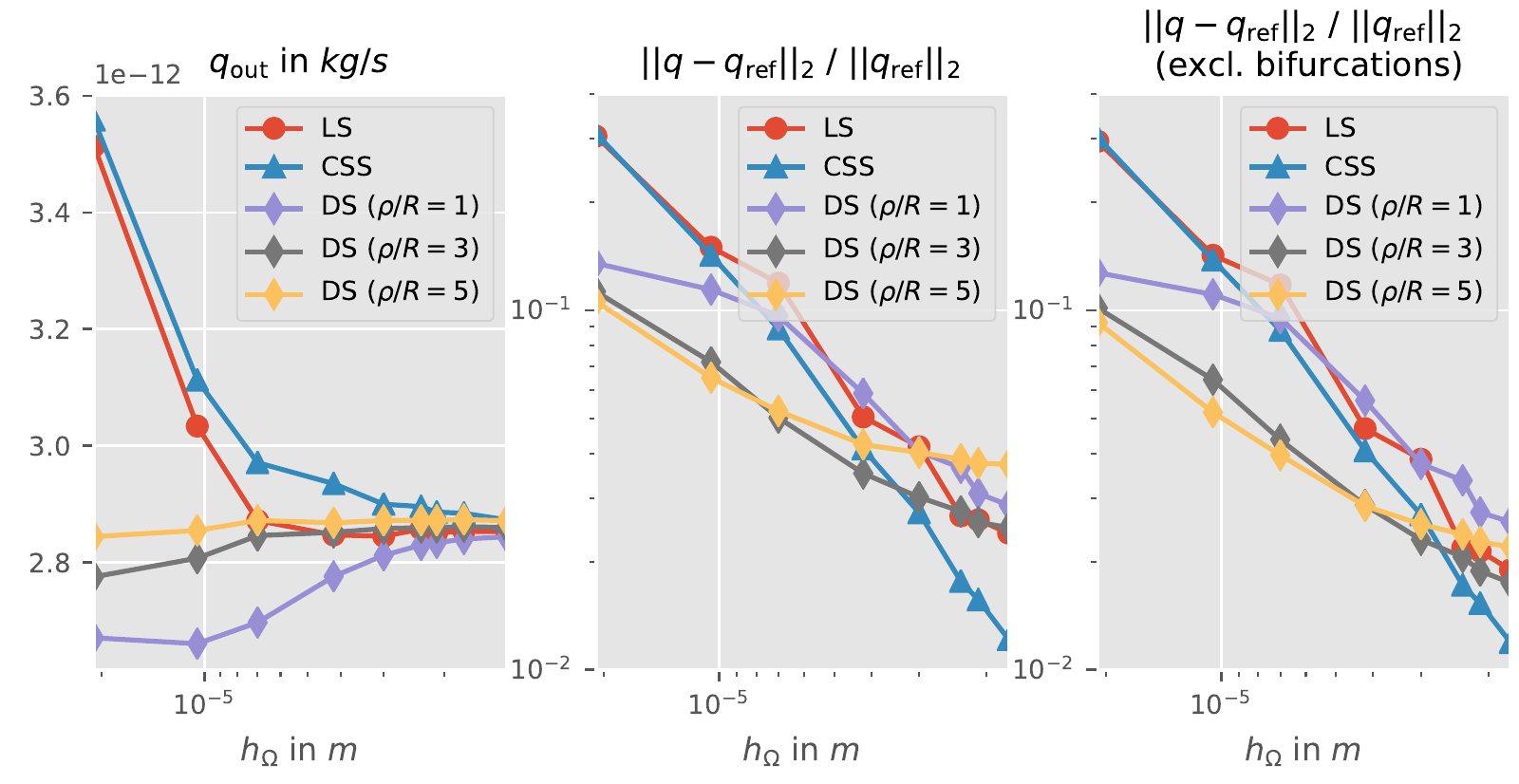}
\caption{Left, the total mass flux emitted by the arterial vessel $q_\text{out}$ for the different methods (\textsc{css}, \textsc{ls}, and \textsc{ds}) with grid refinement.
Center, the relative difference in the $2$-norm of the source term for the different methods and grid refinement, with respect to the reference solution (\textsc{css}).
Right, the relative difference in the $2$-norm of the source term, excluding all cells $K_\Lambda$ that are closer than \SI{10}{\micro\m} to a vessel bifurcation.}
\label{fig:network_results}
\end{figure}

A reference solution is computed using the \textsc{css} method with $h_\Omega = \SI{1.3125}{\micro\m}$ and $h_\Lambda = \SI{0.5}{\micro\m}$.
The network geometry and the $p_v$ reference solution are shown in~\cref{fig:network}.
The source term, $q$, for $h_\Lambda = \SI{0.5}{\micro\m}$, is computed for different $h_\Omega$ using the methods \textsc{css}, \textsc{ls}, and \textsc{ds} with different kernel radii $\varrho/R_i = 1, 3$, and $5$. As in the previous example, we compute the total mass flux, $q_\text{out}$.
The results are shown in~\cref{fig:network_results}.
Firstly, it can be seen that all methods agree well for the total mass flux on the finest grid with maximum differences of about \SI{2}{\percent}.
Secondly, it is evident that the \textsc{ds} approximates $q_\text{out}$ much better for coarser grids, and the approximation gets better with larger kernel support. The \textsc{ds} method with a kernel support radius $\varrho_i = 5 R_i$ very closely approximates $q_\text{out}$ for the
coarsest grid, where $h_\Omega$ is about five times the radius of the largest vessel. If the difference is measured in the relative $2$-norm,
$\Delta q = \vert\vert q-q_\text{ref} \vert\vert_2 / \vert\vert q_\text{ref} \vert\vert_2$, we observe slightly larger differences. On the one hand,
this is due to differences along vessels oscillating around zero, i.e., due to terms that cancel when computing $q_\text{out}$, but not for $\Delta q$.
On the other hand, $\Delta q$ emphasizes larger differences stronger.

\begin{figure}[t!]
\centering
\begin{subfigure}[t]{0.5\textwidth}
    \centering \includegraphics[width=\textwidth]{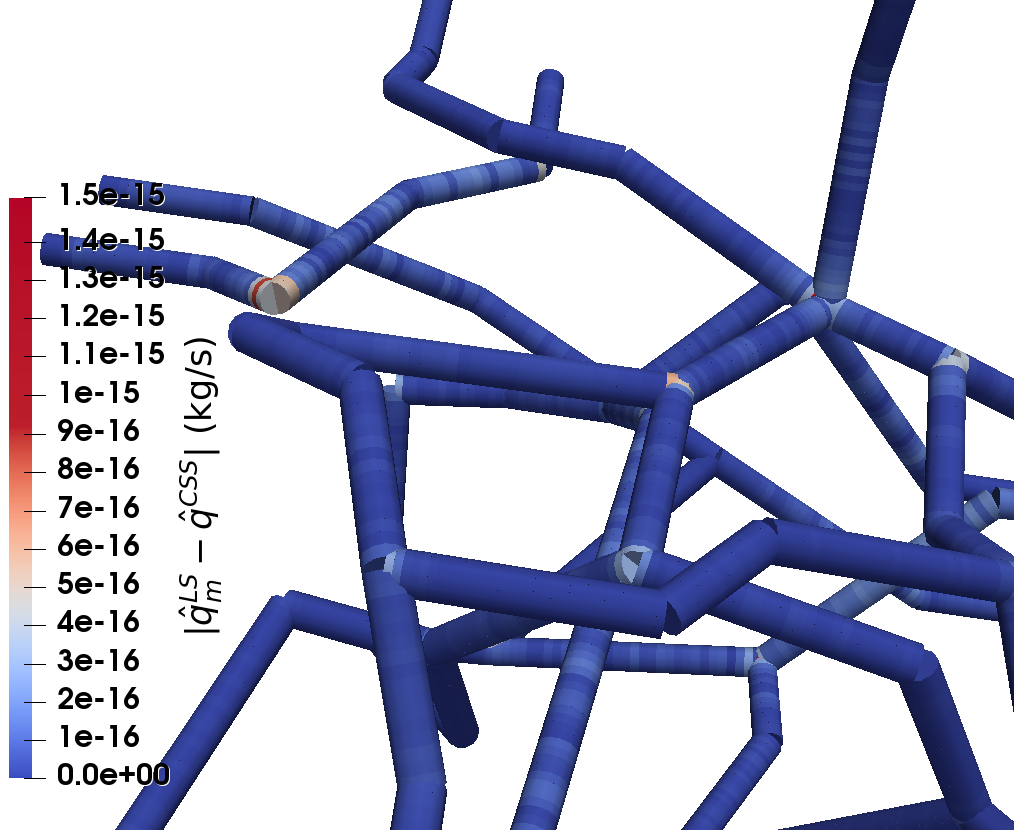}
\end{subfigure}~\begin{subfigure}[t]{0.5\textwidth}
    \centering \includegraphics[width=\textwidth]{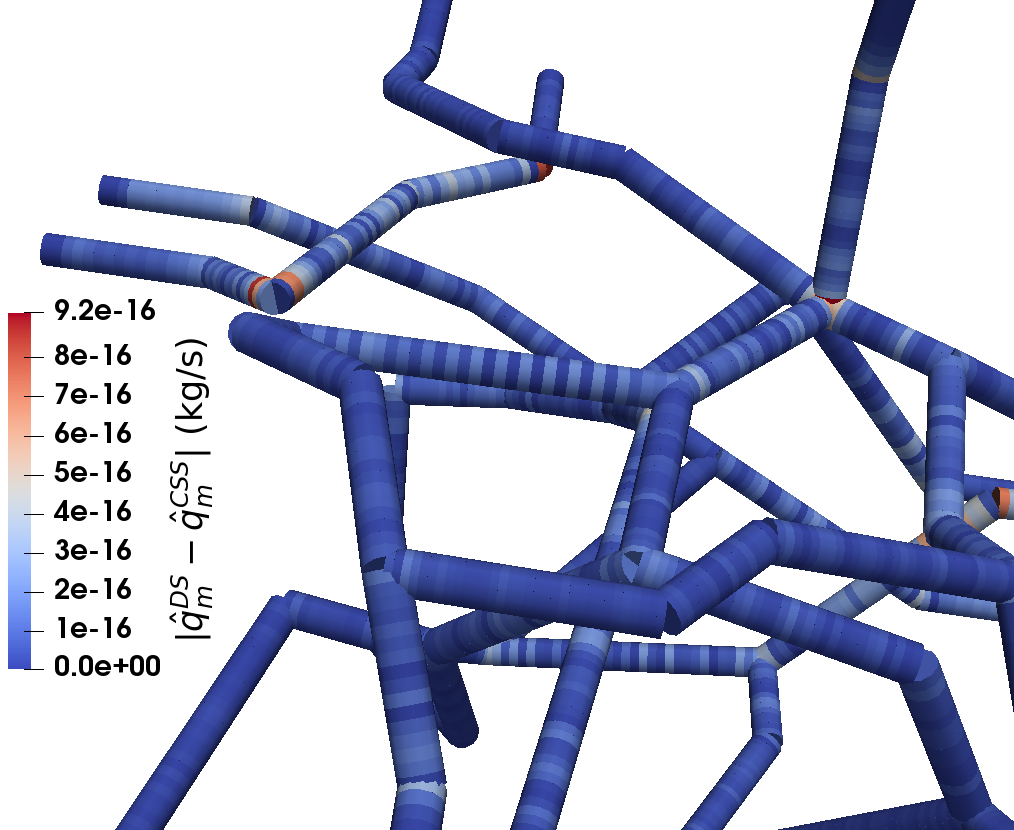}
\end{subfigure}
\caption{Local distribution of differences in the source term for the \textsc{ls} and \textsc{ds} methods with the \textsc{css} method.
         Vessel segments visualized as tubes with a constant radius. The largest differences between the methods can be seen
         in the neighborhood of bifurcations and sharp bends. }
\label{fig:network_localdiff}
\end{figure}

The locality of the difference is visualized in~\cref{fig:network_localdiff},
where we computed the absolute local differences of source terms $\hat{q}_{m,i}$
between the \textsc{ls} and \textsc{ds} methods ($\varrho/R_i = 1$) and the reference solution (with \textsc{css}) for every vessel cell $K_\Lambda$.
The largest differences can be observed in the neighborhood of bifurcations and sharp bends. This difference can be explained by the fact that
the vessels are discretely represented as cylindrical tubes and no explicit intersection geometry is assumed for such features. To the best of our knowledge,
there is currently no mixed-dimension embedded method available that considers particular bifurcation geometries. For the \textsc{ls} method this means that at such features
the mean surface pressure $p_{t,\mathbb{W}}$ contains contributions from the non-physical part of the pressure solution ($r<R$) of the neighboring vessels,
leading to an overestimation of $p_{t,\mathbb{W}}$. This defect is improved in the \textsc{css} method, where, due to the linear continuation of the solution for
$r<R$, the value of $p_{t,\mathbb{W}}$ is closer to the actual pressure on the vessel wall surface. For the \textsc{ds} method, the approximation of
$p_{t,\mathbb{W}}$ is based on an analytical reconstruction from the center-line pressure $p_{t,0}$. However, at bifurcations the kernel support of
the neighboring vessels overlap, which leads to imprecision in the approximation of $\Xi$. \cref{fig:network_results} shows that these effects
lead to differences of \SIrange{3}{4}{\percent} in $\Delta q$ between the methods. Furthermore, it is shown that if the source contributions in
the vicinity of bifurcations ($K_\Lambda$ closer than \SI{10}{\micro\m}) are excluded from the norm, the differences decrease to \SIrange{2}{3}{\percent}.
This shows that the differences are rather local to the bifurcation neighborhood.
Finally, while it cannot be concluded which method is best for the fine grid solutions, we again observe that the \textsc{ds} method, especially the variants
with larger kernel support, better approximate the fine scale solution for large $h_\Omega$ (coarse grid). In fact, the difference to the fine scale solution
(obtained with the \textsc{css} method) for the coarsest grid ($h_\Omega = \SI{21}{\micro\m}$) is \SI{10}{\percent} for the \textsc{ds} method with $\varrho_i = 5R_i$, while it is \SI{30}{\percent} for the \textsc{ls} and \textsc{css} methods.

\FloatBarrier
\section{Summary and conclusion}
\label{sec:discussion}

We presented a new method for modeling tissue perfusion using a mixed-dimension embedded method with distributed sources.
The most prominent difference to existing schemes is that the source term, coupling the vascular and the extra-vascular domains,
is spatially distributed using kernel functions. The mean pressure on the vessel surface is not explicitly computed but locally
reconstructed from the pressure at the vessel center-line using an analytically derived scaling factor. We showed in
four numerical experiment that the result obtained with the new method match well with the results obtained with existing methods.
It was consistently shown in the experiments that the new method converges with a higher rate for the source term $\hat{q}_m$
and the vessel pressure $p_v$ due to an increased regularity in the extra-vascular pressure $p_t$. Furthermore, in all experiments
the new method provided better approximations of the source term $\hat{q}_m$ for coarse extra-vascular grids. In a test with
a realistic vessel network from the rat cortex, using the new method resulted in a three-fold reduction of the error in the source
term for coarse grids in comparison to state-of-the-art methods with respect to a reference solution computed on a fine grid.

In four numerical experiments the new method was compared to existing methods. To this end, all methods have been implemented in
the open source software framework DuMu$^X$~\cite{flemisch2011dumu,Koch2018a}. The implementation effort
and computational costs (at the same grid resolution) of the new method are comparable with existing methods.
However, our results suggest that the new methods can be considered computationally more efficient, since a good approximation
of the source term is already achieved at lower grid resolutions.

The results in this paper suggest that the new method provides the best source term approximations, if the source is distributed
over a volume larger than the vessel itself. In such a case the pressure solution in the extra-vascular space is regularized and thus
may deviate from the physical solution. In turn, the regularized extra-vascular pressure results in better approximations
of source term and vessel pressure. A good approximation of the fluid exchange between vascular and extra-vascular compartments
is crucial in many applications involving transmural transport processes
such as the estimation of contrast agent leakage from the brain microvasculature in multiple sclerosis~\citep{Koch2018b}.
A regularized extra-vascular pressure may be acceptable in cases where it is more interesting how much of a substance
leaves the vascular system rather than its accurate distribution in the extra-vascular space.
Furthermore, it is always acceptable to choose the distribution volume similar to the size of the coarsest neighboring
extra-vascular discretization cell. In such a case, a possible error in the extra-vascular pressure is masked by other
discretization errors. Nevertheless, the description of the source term and thus the vessel pressure may still be improved
significantly by a better reconstruction of the mean vessel surface pressure.

The last numerical experiment revealed that the largest differences between the different mixed-dimension embedded
methods occur in the vicinity of bifurcations. This is due to an imprecise description of bifurcation geometries in the
discrete setting. In order to evaluate methods with respect to the error at bifurcations, and to develop improved
descriptions of the flow around bifurcations, a comparison with methods with spatially resolved interfaces is necessary.
However, our results suggest that discretization errors around bifurcations only affect small parts of the entire system,
such that the current models might be sufficient approximations for most applications.

Finally, due to the improved accuracy at coarse grid resolutions, we consider the new method an important step
towards simulations of larger vessel networks, where fine grid resolutions in the extra-vascular space may get computationally prohibitively expensive.

\section*{Acknowledgements}

This work was financially supported by the German Research Foundation (DGF), within the Cluster of Excellence in Simulation Technology (EXC 310),
and the Collaborative Research Center on Interface-Driven Multi-Field Processes in Porous Media (SFB 1313, Project Number 327154368).

\section*{Appendix}

\subsection*{A1 Domain information for the numerical experiment in \cref{sec:numerical_multiple3d}}

The vessel network used in \cref{sec:numerical_multiple3d} is described in \cref{tab:vesseldata}.

\begin{table}
\centering
\begin{tabular}{|l|l|l||l|l|}
\hline
index & coordinates (\si{\micro\m}) & boundary condition & segment & radius (\si{\micro\m})\\
\hline
0 & (20, 20, -10)  & $v = \SI{1.0}{\milli\meter\per\second}$             & (0, 1) & 3.5 \\
1 & (20, 20, 50)   & -                                                   & (1, 2) & 3.5 \\
2 & (20, 20, 110)  & $\tilde{p}_{v,\text{out}} = \SI{2300}{\pascal}$     & (3, 4) & 3.0 \\
3 & (60, -10, 20)  & $v = \SI{0.8}{\milli\meter\per\second}$             & (4, 5) & 3.0 \\
4 & (60, 20, 20)   & -                                                   & (5, 6) & 3.0 \\
5 & (70, 40, 20)   & -                                                   & (6, 7) & 3.0 \\
6 & (80, 60, 20)   & -                                                   & (8, 9) & 3.0 \\
7 & (110, 60, 20)  & $\tilde{p}_{v,\text{out}} = \SI{3400}{\pascal}$     & (9, 10) & 3.0 \\
8 & (80, 110, 80)  & $v = \SI{0.8}{\milli\meter\per\second}$             & (10, 11) & 3.0 \\
9 & (80, 80, 80)   & -                                                   & (11, 12) & 3.0 \\
10 & (80, 70, 60)  & -                                                   & (13, 14) & 3.0 \\
11 & (20, 80, 20)  & -                                                   & (14, 15) & 3.0 \\
12 & (20, 80, -10) & $\tilde{p}_{v,\text{out}} = \SI{2300}{\pascal}$     & (15, 16) & 3.0 \\
13 & (60, -10, 80) & $\tilde{p}_{v,\text{out}} = \SI{3400}{\pascal}$     & (16, 17) & 3.0 \\
14 & (60, 20, 80)  & -                                                   && \\
15 & (40, 70, 80)  & -                                                   && \\
16 & (20, 80, 80)  & -                                                   && \\
17 & (-10, 80, 80) & $v = \SI{0.8}{\milli\meter\per\second}$             && \\

\hline
\end{tabular}
\caption{Vessel specifications for the numerical experiment in \cref{sec:numerical_multiple3d}.
 The vertex indices and coordinates (in \si{\micro\m}) are listed in the first two columns. Boundary conditions are given, if the vertex is a boundary vertex.
 The 4th and 5th columns list the segment connectivity and the corresponding segment radius (in \si{\micro\m}).}
\label{tab:vesseldata}
\end{table}

\bibliography{kernel}
\bibliographystyle{elsarticle-num}

\end{document}